\documentclass[epj,nopacs]{svjour}
\usepackage{graphics}
\usepackage{amsmath}
\begin{document}
\title{Strain in crystalline core-shell nanowires.}
\author{David Ferrand \inst{1} \and Jo\"{e}l Cibert \inst{2} \thanks {joel.cibert@neel.cnrs.fr}}
\institute{Univ. Grenoble Alpes, Inst NEEL, F-38042 Grenoble,
France\and CNRS, Inst NEEL, F-38042 Grenoble, France}
\date{Received: date / Revised version: date}
\abstract{ We propose a comprehensive description of the strain
configuration induced by the lattice mismatch in a core-shell
nanowire with circular cross-section, taking into account the
crystal anisotropy and the difference in stiffness constants of the
two materials. We use an analytical approach which fully exploits
the symmetry properties of the system. Explicit formulae are given
for nanowires with the wurtzite structure or the zinc-blende
structure with the hexagonal / trigonal axis along the nanowire, and
the results are compared to available numerical calculations and
experimental data on nanowires made of different III-V and II-VI
semiconductors. The method is also applied to multishell nanowires,
and to core-shell nanowires grown along the $<001>$ axis of cubic
semiconductors. It can be extended to other orientations and other
crystal structures.
} %end of abstract
\maketitle
\section{Introduction}

Semiconductor nanowires (NWs) are often grown in the form of
core-shell structures, in order to achieve better photonic and
electronic properties: the active core is isolated from the surface
defects and traps in order to obtain a better luminescence
efficiency, sharper linewidths, longer coherence times and higher
mobility, or even a better chemical stability. As the lattice
parameter of the shell is generally different from that of the core,
and since coherent structures are contemplated with no misfit
defects at the interface, the elastic strain induced in the core and
its effect on the electronic properties have to be taken into
account. In turn, the built-in strain can be used as a further
adjustable parameter: strain engineering can be used to lower the
degeneracy in the valence band and select the type of holes with a
larger spin for spintronics applications (for instance a larger
spin-carrier coupling in diluted magnetic semiconductors)
\cite{Wojn12}, or a smaller longitudinal mass to achieve a better
mobility in transport properties \cite{Niquet}. The strain can also
be designed to induce a built-in piezoelectric field, resulting in a
faster separation of the electron-hole pairs in photovoltaic
applications \cite{BoxbergNL}. Finally, strain is an important
parameter when engineering Si-Ge NWs to obtain direct bandgap
configurations and efficient emission of light \cite{Amato}.

Analytical expressions exist for a core-shell structure made of
elastically isotropic materials \cite{Gutkin00}. However, the
crystal structure results in anisotropic elastic properties, the
core and shell materials have different values of the stiffness
constants, and the NW shape can deviate from the ideal cylinder with
a circular base and for instance feature facetting. As a result,
calculating the strain configuration in a real semiconductor
core-shell NW is not an easy task: quantitative descriptions usually
imply to compute numerically the local strain, either using a
microscopic model such as the valence force field model, or
performing a finite element treatment of the continuum elasticity
theory \cite{Gronqvist}. Nevertheless, an analytical description,
such as what has been developed and reviewed in
Ref.~\cite{Maranganti} for the case of quantum dots and NWs embedded
in an infinite or semi-infinite material, remains the best starting
point for an implementation of the strain-related mechanisms
governing the electronic properties, through deformation potentials
and piezoelectric fields.

A quantitative, fully analytical solution for core-shell NWs, taking
into account the crystal structure, can be found, and this is the
purpose of the present study. Starting with the well known
expression for isotropic materials (and their extension for the
transversely isotropic materials), we propose solutions for the most
often encountered cases of zinc-blende and wurtzite semiconductors.
We give analytical expressions for the strain in the core and in the
shell, and for their effect on the extrema of bands, and we compare
these predictions to the results of microscopic calculations and
experimental data.

In most cases, deviations from the cylindrical strain configuration
are found. In two typical cases (with zinc-blende or diamond
semiconductor NWs along $<111>$ and along $<001>$), we identify the
resulting strain configuration to first-order in the parameter
describing the cubic anisotropy and we show that these deviations
from cylindrical symmetry are rather small. These two cases
illustrate two non-isotropic strain configurations: warping along
the NW axis, and anisotropy of the in-plane strain. Other
orientations and crystal structures are expected to feature a
combination of these two configurations.

To sum up our results: (\emph{i}) the cylindrical approximation is
surprisingly good, provided one uses the appropriate truncation of
the stiffness tensor which is given here; (\emph{ii}) the result is
exact for wurtzite NWs grown along the hexagonal axis; (\emph{iii})
a first-order treatment of the anisotropy quantitatively agrees with
available numerical results for zinc-blende NWs; the additional
strain components are negligible in the core but they take
significant values in the shell; (\emph{iii}) the present method is
readily extended to other structures or orientations, and to
multishell NWs.

The paper is organized as follows: section II is a short summary of
the problem to be solved and of results which are well-known for
isotropic materials. In section III, we obtain analytical
expressions of the strain configuration in wurtzite semiconductor
NWs grown along the hexagonal axis; the transfer matrix approach
allows us to consider both core-shell and multishell NWs. In
sections IV and V, we use a perturbation method to describe the more
complex strain configuration present in zinc-blende semiconductor
NWs grown along the trigonal axis and along the cubic axis.
\section{Strain and electrons in a core-shell NW}
\subsection{The displacement field in an infinite core-shell NW}
In this section we recall the well-known strain configuration in an
infinitely long core-shell NW with circular cross-section, made of
isotropic materials, in order to identify and illustrate the effects
of the two elements of symmetry on the displacement field and the
Lam\'{e}-Clapeyron-Navier equation. We consider a cylinder-shaped
core (superscript or subscript $c$), infinitely long, with a
circular cross section of radius $r_c$, embedded in a shell
(superscript or subscript $s$) of radius $r_s$. We note $z$ the NW
axis, ($r$, $\theta$) or ($x$, $y$) the in-plane coordinates
measured from the NW axis. The two materials have the same crystal
structure and the same orientation, with different values of the
lattice constants $a_s$ and $a_c$. The growth is assumed to be
coherent, with no misfit defect at the interface, so that the
lattice mismatch $f=(a_s-a_c)/a_c$ is fully accommodated by elastic
strain.

The general solution involves calculating the displacement field
$\textbf{u}(\textbf{r})$ which relates the position of any point
$\textbf{r}$ in the strained material to its value in the
mismatched, unstrained system. The local deformation, in the
vicinity of a point $\textbf{r}$, is fully described by the tensor
of the derivatives of $\textbf{u}(\textbf{r})$, $\partial
u_i/\partial x_j$: the symmetric part is the strain tensor,
$\varepsilon_{ij}=\frac{1}{2}(\partial u_i/\partial x_j+\partial
u_j/\partial x_i)$, associated to elastic energy, while the
antisymmetric part $\frac{1}{2}(\partial u_i/\partial x_j-\partial
u_j/\partial x_i)$ describes a local rotation. In the presence of
body forces per unit volume $\textbf{F}(\textbf{r})$, the
equilibrium condition, $\sum_j \partial \sigma_{ij}/\partial
x_j+F_i=0$, can be expressed as the Lam\'{e} - Clapeyron - Navier
equation (there is one equation for each value of $i$ and
$x_i=x,y,z$),
\begin{eqnarray} \label{LCNgeneral}
\label{eq1} \sum_{jkl} c_{ijkl} \frac{\partial}{\partial x_j}
(\frac{\partial u_k}{\partial x_l}+\frac{\partial u_l}{\partial
x_k})+F_i=0
\end{eqnarray}
In this equation, the $c_{ijkl}$ are the components of the stiffness
tensor, which relates the stress tensor $\sigma_{ij}$ to the strain
tensor $\varepsilon_{kl}$ through the Hooke's law,
$\sigma_{ij}=\sum_{kl} c_{ijkl}\varepsilon_{kl}$. The number of
independent components $c_{ijkl}$ is determined by the symmetry
properties of the material \cite{Nye}.

In a core-shell NW, we apply the Lam\'{e} - Clapeyron - Navier
equation within each constituent; the body forces are zero, but we
have to apply proper boundary conditions \cite{Tsukrov} at the
surface and at the interface. A first series of conditions ensure
the stability of the interface/surface: stress components applied to
the surface ($\sigma_{rr}$, $\sigma_{r\theta}$ and $\sigma_{rz}$)
vanish, and they are equal on both sides of the interface.
Additional conditions state the continuity of the lattice: the
displacement field $\textbf{u}(\textbf{r})$ must compensate for the
lattice mismatch $f$. All these conditions are actually the same as
for a thin epitaxial layer, but then the condition on the continuity
of the lattice can be expressed on the in-plane strain components
\cite{Marcus}.

In addition, for an infinitely long NW, the overall translational
invariance along the axis must be maintained (and it is known also
that in a NW of finite length, according to the Saint-Venant
principle, this holds everywhere but for a segment of length equal
to about the diameter at each end). Translational invariance means
that the relative displacement of two neighboring points is
independent of $z$,\emph{ i.e.}, that all derivatives of
$\textbf{u}(\textbf{r})$ are independent of $z$: $\partial/\partial
z~(\partial u_i/\partial x_j)=0$, or $\partial/\partial
x_j~(\partial u_i/\partial z)=0$, hence $\partial u_i/\partial z$ is
a constant $C_i$ independent of $\textbf{r}$, $\partial u_i/\partial
z=C_i$, and $u_i(\textbf{r})=C_iz+D+u_i(x,y)$. Note that the $C_xz$,
$C_yz$ and $u_z(x,y)$ contributions correspond to shear strains
($\varepsilon_{xz}$, $\varepsilon_{yz}$) and are often excluded by
symmetry. Finally, the equilibrium with respect to a translation
along the NW axis requires that the longitudinal stress integrated
over the NW section be zero.

Once determined the displacement field $\textbf{u}(\textbf{r})$
obeying the Lam\'{e} - Clapeyron - Navier equation and the boundary
conditions, the strain tensor can be introduced into the so-called
deformation potentials \cite{deformation} and the possible
piezoelectric field is calculated; the positions of the conduction
and valence band edges follow.
\subsection{The simple case of elastically isotropic materials}
The solution for an infinitely long, circular core-shell structure
made of elastically isotropic materials, is well known
\cite{Gutkin00}. We briefly recall the main results, our goal being
to examine what will remain valid if materials with a lower symmetry
are involved.

The Lam\'{e} - Clapeyron - Navier equation writes
\begin{equation}\label{LCNiso}
    \mu \sum_j \frac{\partial ^2 u_i}{\partial  x_j~^2} +(\lambda + \mu)
\sum_j \frac{\partial ^2 u_j}{\partial x_i \partial x_j}=0
\end{equation}

which contains three equations, for $x_i=x$, $y$ and $z$,
respectively. A more compact form better evidences the spherical
symmetry:
\begin{equation}\label{LCNvector}
\mu \Delta\textbf{u} + (\lambda + \mu) \nabla (\nabla .
\textbf{u})=\textbf{0}
\end{equation}
Here the stiffness tensor has only two independent components: the
so-called Lam\'{e} coefficients $\mu=c_{ijij}=c_{ijji}$ and
$\lambda=c_{iijj}$ for $i \neq j$, with $c_{iiii}=\lambda + 2 \mu$.
All other components vanish.

If we omit the terms which vanish due to the invariance by
translation or would correspond to axial shear strains (according to
the discussion in the previous section), the Lam\'{e} - Clapeyron -
Navier equation restricts to:
\begin{eqnarray}\label{LCNtranslation}
&&\mu (\frac{\partial^2}{\partial y^2}+\frac{\partial^2}{\partial
x^2})u_x+ (\lambda + \mu) \frac{\partial}{\partial x}(\frac{\partial
u_x}{\partial x}+\frac{\partial u_y}{\partial y})=0\nonumber\\ &&\mu
(\frac{\partial^2}{\partial y^2}+\frac{\partial^2}{\partial
x^2})u_y+ (\lambda + \mu)
\frac{\partial}{\partial y}(\frac{\partial u_x}{\partial x}+\frac{\partial u_y}{\partial y})=0\nonumber\\
&&\mu (\frac{\partial^2}{\partial y^2}+\frac{\partial^2}{\partial
x^2})u_z=0
\end{eqnarray}

\begin{figure}
\resizebox{0.5\textwidth}{!}{\includegraphics {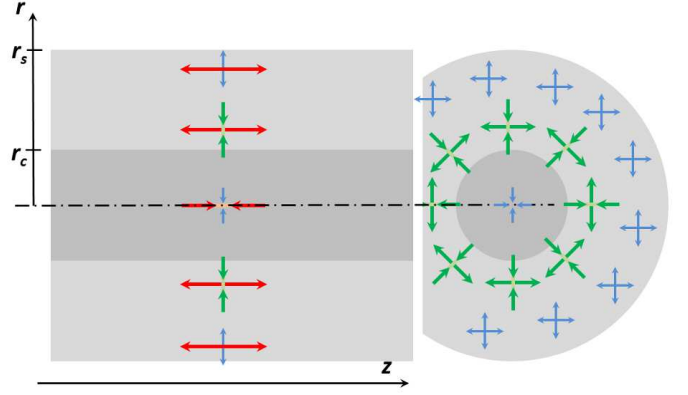}
}\caption{(color online) Strain distribution in a cylindrical
core-shell NW. Arrows indicate the longitudinal strain (in red), the
inhomogeneous shear strain in the shell (in green), and the rest of
the strain - uniform and isotropic in the plane (in blue). The
lattice parameter is assumed to be smaller in the shell than in the
core ($f<0$). \label{strainIso}}
\end{figure}

As the strained system obviously retains the cylindrical symmetry,
we write the displacement field in cylindrical coordinates, keeping
only the relevant variables: $u_r(r)$, $u_\theta=0$, with
$u_x=u_r(r) \cos \theta $ and $u_y=u_r(r) \sin \theta$. An in-plane
dependence of $u_z$ would imply shear strain components
$\varepsilon_{rz}$ which are excluded, hence $u_z(z)=Cz+D$. Finally
the Lam\'{e} - Clapeyron - Navier equation is reduced to equating to
zero the Laplacian of the in-plane displacement, hence
$d^2u_r/dr^2+du_r/rdr-u_r/r^2=0$, and $u_r(r)=Ar+Br_c^2/r$, with
parameters $A$, $B$, $C$ and $D$ to be determined in each material.
The non-vanishing components of the strain tensor are thus the
longitudinal expansion $\varepsilon_{zz}=du_z/dz=C$, the radial
expansion $\varepsilon_{rr}=du_r/dr=A-B r_c^2/r^2$, and the angular
expansion $\varepsilon_{\theta \theta}=u_r/r=A+Br_c^2/r^2$. Note
that $B$ vanishes in the core (to avoid diverging terms at the axis,
$r=0$); also, $D$ represents a global displacement of the core or
the shell, hence $D=0$. As a result, see fig.~\ref{strainIso}, the
strain (and the stress) are uniform in the core; in the shell, there
is also a uniform component, and a non uniform shear component,
rotating around the interface and close to it. Note that the stress
component $\sigma_{zz}^c$ is uniform also in the shell (the
non-uniform $B r_c^2/r^2$ terms in $\varepsilon_{\theta\theta}^s$
and $\varepsilon_{rr}^s$ cancel each other when applying the Hooke's
law).

The two parameters $A_c$ and $C_c$ in the core, and the three
parameters $A_s$, $B_s$ and $C_s$ in the shell, are determined from
the boundary conditions. At the interface, the matching along $z$
(written on $u_z$ or $\varepsilon_{zz}$) implies $C^c-C^s=f_{\|}$,
and the matching in the plane is realized simultaneously on $u_r$
and $\varepsilon_{\theta \theta}$ if $A^c-A^s-B^s=f_{\bot}$. We
identify the mismatch $f_{\|}$ in the direction of the NW axis, and
the mismatch $f_{\bot}$ in the plane perpendicular to the axis:
although this is not done usually - and not needed for isotropic
materials - that will allow a better understanding of the result.
The stress components are such that
$\sigma_{rr}^c(r_c)-\sigma_{rr}^s(r_c)=0$ at the interface and
$\sigma_{rr}^s(r_s)=0$ at the sidewall. The other components
($\sigma_{r \theta}$ and $\sigma_{rz}$) automatically vanish. The
longitudinal stress integrated over the NW section vanishes: as both
$\sigma_{zz}^s$ and $\sigma_{zz}^c$ are uniform, the condition is
simply $\eta \sigma_{zz}^c + (1-\eta) \sigma_{zz}^s=0 $ where $\eta$
is the ratio of the core to NW cross-section areas (for a NW with
circular cross-section, $\eta=r_c^2 / r_s^2$).

A straightforward calculation then gives the complete set of strain
components
\begin{eqnarray} \label{StrainSameStiffness}
\varepsilon_{zz}^c&=&(1-\eta )f_{\|}  \nonumber \\
\varepsilon_{zz}^s&=&-\eta f_{\|} \nonumber \\
\frac{\varepsilon_{\theta \theta}^s -\varepsilon_{rr}^s}{2} &=& B_s
\frac{r_c^2}{r^2}\nonumber\\
\varepsilon_{\theta \theta}^c =\varepsilon_{rr}^c&=&(1-\eta )
(f_{\bot}+B_s)
 \nonumber\\
\frac{\varepsilon_{\theta \theta}^s +\varepsilon_{rr}^s}{2} &=&-
\eta (f_{\bot}+B_s)
 \end{eqnarray}
where
\begin{eqnarray} \label{strainiso}
B_s&=&-\frac{2 (\lambda+\mu) f_{\bot} + \lambda f_{\|}}{2 (\lambda + 2 \mu)} \nonumber\\
(f_{\bot}+B_s)&=&\frac{2 \mu f_{\bot}-\lambda f_{\|} }{2(\lambda + 2 \mu)} \nonumber\\
\end{eqnarray}

The longitudinal strain $\varepsilon_{zz}$ (red arrows in
fig.~\ref{strainIso}) results from the lattice mismatch in the
direction of the NW axis, which is shared between the core and the
shell with a weight inversely proportional to their area (in a way
similar to the strain distribution in a free-standing superlattice,
where the lattice mismatch is shared with a weight inversely
proportional to the thickness of each layer). A narrow core is fully
strained to the thick shell (and a thin shell to a wide core). The
main part of the in-plane lattice mismatch is accommodated by the
shear strain in the shell rotating around the interface (green
arrows in fig.~\ref{strainIso}). The rest of the in-plane strain
consists in a uniform in-plane strain in the core and a uniform
component in the shell (blue arrows in fig.~\ref{strainIso}): these
components result from the competition between a direct effect of
the lattice mismatch in the plane, and an indirect effect of the
longitudinal strain. As a result, they can be quite small.

For a thin shell, $\eta=1$, the core is unstrained, and the shell
strain writes $\varepsilon_{zz}^s=\varepsilon_{\theta \theta}^s=-f$,
$\varepsilon_{rr}^s=[2\lambda/(\lambda+2\mu)]f$, which is the result
for a thin epitaxial layer on a plane substrate.

The previous result can be extended \cite{Aifantis} to the case of
two isotropic materials with different values of the shear modulus,
but the same value of the Poisson ratio. In terms of Lam\'{e}
coefficients, that means $\lambda_s/\lambda_c=\mu_s/\mu_c$. Complete
expressions of the stress tensor are given in ref. \cite{Aifantis}.
We will generalize these expressions in the following section taking
into account the crystal structure.
\subsection{The effect on the electronic properties}
Two mechanisms affecting the electronic properties of a core-shell
NW are determined by the strain configuration.
\begin{itemize}
  \item There is a direct effect of strain on the bands of a semiconductor;
around the band edges, it is described phenomenologically by the
so-called deformation potentials. For instance, in a zinc-blende
semiconductor, the isotropic strain (change of volume),
$(\varepsilon_{xx}+\varepsilon_{yy}+\varepsilon_{zz})$, induces a
shift of the conduction band and an average shift of the valence
band at the center of the Brillouin zone. A shear strain, such as
$(2\varepsilon_{zz}-\varepsilon_{xx}-\varepsilon_{yy})$ induces a
splitting of the valence band edge.
  \item When NWs are grown along a polar axis, they are expected to present
a polarization due to the piezoelectric effect. This is the case of
NWs with the wurtzite structure grown along the $c$-axis, as well as
NWs with the zinc-blende structure grown along the $<111>$ axis. The
relevant strain components entering the longitudinal polarization
are \cite{BoxbergNL} $\varepsilon_{zz}$ and
$(\varepsilon_{rr}+\varepsilon_{\theta\theta})$ in the first case,
and $(2\varepsilon_{zz}-\varepsilon_{xx}-\varepsilon_{yy})$ in the
second case.
\end{itemize}
In addition, confinement effects should be taken into account if the
NW radius is small enough, and the confining potential is modified
by these two mechanisms.

It is interesting to compare the results for a thin core and that
for a thin epitaxial layer, both considered as the active medium of
the structure. In both cases there is an isotropic strain and a
shear strain. The isotropic strain is
$(\varepsilon_{xx}+\varepsilon_{yy}+\varepsilon_{zz})$ or
$(\varepsilon_{rr}+\varepsilon_{\theta\theta}+\varepsilon_{zz})=[4\mu/(\lambda+2\mu)]f$
in both cases. The shear strain is
$(2\varepsilon_{zz}-\varepsilon_{xx}-\varepsilon_{yy})=-2[(
3\lambda+2\mu)/(\lambda+2\mu)]f$ in the thin epitaxial layer, and
$(2\varepsilon_{zz}-\varepsilon_{rr}-\varepsilon_{\theta\theta})=[(3\lambda+2\mu)/(\lambda+2\mu)]f$.
The same result holds for a thick core, with $f$ replaced by $\eta
f$. Hence the ratio of the valence band splitting to the shift is
(1) of opposite sign, and (2) twice smaller, in the core of a NW
than in an epitaxial layer made of elastically isotropic materials.
This property will be checked below in the presence of crystalline
anisotropy: we will show that the factor is not exactly 2.
\subsection{Crystalline semiconductor NWs}
Our goal is to take into account the crystal structure of the
semiconductors, by using the stiffness tensor with the appropriate
symmetry. We will consider explicitly three cases: hexagonal
(wurtzite) structure with the NW axis along the $c$-axis, and cubic
(zinc-blende or diamond) structure with the NW axis along $<001>$ or
$<111>$. We ignore facetting and consider a NW with a circular
cylinder shape. We will show that
\begin{itemize}
  \item In the case of a wurtzite NW grown along the six-fold axis, the transversely isotropic solution is exact.
  \item In the case of a zinc-blende NW grown along a trigonal axis,
  the transversely isotropic solution is an excellent approximation, which reproduces quantitatively
  the results of numerical approaches. Deviations due to the cubic
  anisotropy appear in the form of a warping along the axis, of three-fold symmetry, and can
  be found as the response of an elastically isotropic system to a
  distribution of body forces parallel to the NW axis.
  \item In the case of a zinc-blende NW grown along a tetragonal
  axis, a transversely isotropic approximation is proposed.
  Deviations with four-fold symmetry are found and calculated as the response to a distribution of body
  forces perpendicular to the NW axis.
\end{itemize}

The stiffness tensor is written using the Voigt notation,
$\varepsilon_1=\varepsilon_{xx}$,...,
$\varepsilon_4=\varepsilon_{yz}+\varepsilon_{zy}$,..., and
$c_{xxxx}=c_{11}$, $c_{zzzz}=c_{33}$, $c_{yzyz}=c_{44}$,
$c_{xyxy}=c_{66}$ and so on.

The stiffness tensor for the zinc-blende structure reflects the
cubic symmetry \cite{Nye}. It contains three independent terms and
the Voigt notation in the cubic axes is:

\begin{equation}\label{stiffnessZB}
\begin{pmatrix}
c_{11}&c_{12}&c_{12}&0&0&0 \\
c_{12}&c_{11}&c_{12}&0&0&0 \\
c_{12}&c_{12}&c_{11}&0&0&0 \\
0&0&0&c_{44}&0&0 \\
0&0&0&0&c_{44}&0 \\
0&0&0&0&0&c_{44} \\
\end{pmatrix}
\end{equation}

The anisotropy is characterized by the parameter
$c=(c_{11}-c_{12}-2c_{44})$. If $c=0$, the energy of a tetragonal
shear strain (characterized by $c_{11}-c_{12}$) equals that of a
trigonal shear strain (characterized by $2c_{44}$) and the spherical
symmetry is restored. Usual semiconductors have $c<0$: they are
harder against a trigonal stress, which directly involves a change
of bond length, than against a tetragonal stress which is
accommodated mainly by bond rotation. As a result, they are harder
along a $<111>$ direction and softer along a $<001>$ direction, with
$<110>$ in between \cite{Wortman}.

In the wurtzite structure, with $z$ along the $c$-axis and $x$, $y$
in the perpendicular plane, symmetry considerations imply identities
such as $c_{22}=c_{11}$ or $c_{66}=(c_{11}-c_{12})/2$, so that the
stiffness tensor has five independent components \cite{Nye}:

\begin{equation}\label{stiffnessW}
\begin{pmatrix}
c_{11}&c_{12}&c_{13}&0&0&0 \\
c_{12}&c_{11}&c_{13}&0&0&0 \\
c_{13}&c_{13}&c_{33}&0&0&0 \\
0&0&0&c_{44}&0&0 \\
0&0&0&0&c_{44}&0 \\
0&0&0&0&0&\frac{c_{11}-c_{12}}{2} \\
\end{pmatrix}
\end{equation}

It is invariant under any rotation around the $c$-axis.
\section{Hexagonal semiconductors along the c-axis}
We consider a NW with the wurtzite structure, and its axis parallel
to the $c$ axis. We take the $z$ axis along this axis, and $x$ and
$y$ two arbitrary axes in the basal plane. Note that the lattice
mismatch along the $c$-axis, $f_{\|}$, and perpendicular to it,
$f_{\bot}$, may be different.
\subsection{Calculation}
Our calculation is similar to that of Ref.~\cite{Warwick}, where the
stress is calculated for coaxial cylinders with transverse isotropy:
indeed this is the case for hexagonal semiconductors around the
c-axis. In this section, we give the full expressions of the strain,
which are the useful parameters to calculate the local potential and
the piezoelectric field. Moreover, this part constitutes our first
step for the calculation of strain in systems lacking transverse
isotropy.

The complete Lam\'{e} - Clapeyron - Navier equation
(Eq.~\ref{LCNgeneral}) is written in Appendix \ref{completeLCN}
(Eq.~\ref{LCNhexafull}). Omitting terms which vanish due to
invariance by translation, we obtain:
\begin{eqnarray}\label{LCNhexacut}
&&\frac{c_{11}-c_{12}}{2} (\frac{\partial^2}{\partial
y^2}+\frac{\partial^2}{\partial x^2})u_x+ \frac{c_{11}+c_{12}}{2}
\frac{\partial}{\partial x}(\frac{\partial u_x}{\partial x}+\frac{\partial u_y}{\partial y})=0\nonumber\\
&&\frac{c_{11}-c_{12}}{2} (\frac{\partial^2}{\partial
y^2}+\frac{\partial^2}{\partial x^2})u_y+ \frac{c_{11}+c_{12}}{2}
\frac{\partial}{\partial y}(\frac{\partial u_x}{\partial x}+\frac{\partial u_y}{\partial y})=0\nonumber\\
&&(\frac{\partial^2}{\partial y^2}+\frac{\partial^2}{\partial
x^2})u_z=0
\end{eqnarray}

It reproduces exactly the Lam\'{e} - Clapeyron - Navier equation of
an elastically isotropic material, Eq.~\ref{LCNtranslation}. Hence,
as the boundary conditions are invariant under a rotation around the
NW axis (this is due to the invariance of the stiffness tensor noted
above), the general solution for Eq.~\ref{LCNhexacut} is the same as
that of Eq.~\ref{LCNtranslation}, $u_z(z)=Cz$ and
$u_r(r)=Ar+Br_c^2/r$. Furthermore, as this solution is such that the
terms of Eq.~\ref{LCNhexafull} omitted in Eq.~\ref{LCNhexacut} all
vanish, it is the exact solution of the complete equation,
Eq.~\ref{LCNhexafull}.

Applying the Hooke's law to the boundary conditions of a core-shell
NW as in the previous section (at the interface, step in
$u_z/z=\varepsilon_{zz}$ and in $u_r/r=\varepsilon_{\theta \theta}$
to accommodate the lattice mismatch $f_{\|}$ and $f_{\bot}$ with no
misfit dislocation, and equilibrium of $\sigma_{rr}$; at the
sidewall, $\sigma_{rr}^s(r_s)=0$; along the $z$-axis, $\eta
\sigma_{zz}^c+(1-\eta)\sigma_{zz}^s=0$), we obtain the strain tensor
by inverting a system of linear equations:

\begin{eqnarray}\label{fullHexa}
&&\begin{pmatrix}
0&0&0&1&-1 \\
1&-1&-1&0&0 \\
(c_{11}^c+c_{12}^c)&-(c_{11}^s+c_{12}^s)&(c_{11}^s-c_{12}^s)&c_{13}^c&-c_{13}^s \\
\eta 2c_{13}^c&(1-\eta)2c_{13}^s&0&\eta c_{33}^c&(1-\eta)c_{33}^s \\
0&(c_{11}^s+c_{12}^s)&-\eta (c_{11}^s-c_{12}^s)&0&c_{13}^s \\
\end{pmatrix}\nonumber\\
&&\times
\begin{pmatrix}
A^c\\
A^s\\
B^s\\
C^c\\
C^s\\
\end{pmatrix}
=
\begin{pmatrix}
f_{\|}\\
f_{\bot}\\
0\\
0\\
0\\
\end{pmatrix}
\end{eqnarray}

A direct numerical calculation is possible, however it is
interesting to write the boundary conditions using a transfer matrix
method, which can be generalized to multishell NWs \cite{Tsukrov}:
then, we have to solve a system of two linear equations, instead of
5 for a core-shell NW and $(3n+2)$ for a NW with $(n-1)$ shells.

\subsubsection {Transfer matrix}

We thus consider a multishell NW made of a core of radius $r_0$, and
several layers of radius $r_i$ and lattice mismatch $f_{\| i}$ and
$f_{\bot i}$ with respect to the core material, with a uniform
stiffness tensor over the whole NW. The radius of the last layer,
$i=s$, is the NW radius. The relative cross section area of each
layer is $\eta_i=(r_i^2-r_{i-1}^2)/(r_s^2)$.

Within each material, we define a matrix $\textsf{\textbf{M}}(\rho)$
relating the relevant components of displacement and stress to the
$A$, $B$, $C$ parameters, with $\rho=r/r_0$:
\begin{equation}\label{transfer}
    \begin{pmatrix}
    \frac{u_z}{z}\\
    \frac{u_r}{r}\\
    \sigma_{rr}
    \end{pmatrix}
    =
    \textsf{\textbf{M}}(\rho)
    \begin{pmatrix}
    C\\
    A\\
    B
    \end{pmatrix} \nonumber
\end{equation}
with
\begin{eqnarray}\label{}
    \textsf{\textbf{M}}(\rho)&=&
    \begin{pmatrix}
    1&0&0\\
    0&1&\rho^{-2}\\
    c_{13}&(c_{11}+c_{12})&-(c_{11}-c_{12})\rho^{-2}
    \end{pmatrix}
   \nonumber\\
     & =& \textsf{\textbf{M}}(1)
   \begin{pmatrix}
    1&0&0\\
    0&1&0\\
    0&0&\rho^{-2}
    \end{pmatrix}
   \nonumber
\end{eqnarray}
and
\begin{equation}\label{}
     \textsf{\textbf{M}}^{-1}(\rho)=\frac{1}{2c_{11}}
    \begin{pmatrix}
    2c_{11}&0&0\\
    -c_{13}&(c_{11}-c_{12})&1\\
    c_{13}\rho^2&(c_{11}+c_{12})\rho^2&-\rho^2
    \end{pmatrix}   \nonumber
\end{equation}

In the general case, the values of the stiffness constants are
specific to the material which makes the layer $i$, and accordingly
there is a matrix $\textsf{\textbf{M}}_i$ appropriate to each
material.

The boundary condition at the interface between layers $i$ and
$(i+1)$, at $\rho = \rho_i$, is
\begin{equation} \label{interface}
    \textsf{\textbf{M}}_{i+1}(\rho_i)
\begin{pmatrix}
    C_{i+1}\\
    A_{i+1}\\
    B_{i+1}
    \end{pmatrix}
    =
    \textsf{\textbf{M}}_i(\rho_i)
\begin{pmatrix}
    C_i\\
    A_i\\
    B_i
    \end{pmatrix}
    -
\begin{pmatrix}
    f_{\|(i+1)}\\
    f_{\bot (i+1)}\\
    0
    \end{pmatrix}
    +
    \begin{pmatrix}
    f_{\|i}\\
    f_{\bot}i\\
    0
    \end{pmatrix}
    \end{equation}
This condition can be re-written
\begin{eqnarray}\label{recurrence}
\begin{pmatrix}
    C_{i+1}\\
    A_{i+1}\\
    B_{i+1}
    \end{pmatrix}
    &=&
 \textsf{\textbf{M}}^{-1}_{i+1}(\rho_i) \textsf{\textbf{M}}_i(\rho_i)
 \begin{pmatrix}
    C_i\\
    A_i\\
    B_i
    \end{pmatrix}
    \nonumber\\
    &&-    \textsf{\textbf{M}}_{i+1}^{-1}(\rho_i)
\left[
\begin{pmatrix}
    f_{\|(i+1)}\\
    f_{\bot (i+1)}\\
    0
    \end{pmatrix}
    -
    \begin{pmatrix}
    f_{\|i}\\
    f_{\bot}i\\
    0
    \end{pmatrix} \right]
    \end{eqnarray}

Eq.~\ref{recurrence} establishes a relation of recurrence from the
parameters on the inner side of the interface, to those on the outer
side. Repeating Eq.~\ref{recurrence} from shell to shell, we obtain
the set of parameters ($C_i$, $A_i$, $B_i$) as a function of those
of the core, ($C_0$, $A_0$, with $B_0$=0). The two core parameters
are finally determined by the two boundary conditions on the stress
at the sidewall and along $z$.

The first boundary conditions, $\sigma_{rr}=0$ at the sidewall, is
written using the projection
$^t\textsf{\textbf{P}}_r=\begin{pmatrix}
    0&
    0&
    1
    \end{pmatrix}$,
\begin{equation}\label{rrcondition}
    ^t\textsf{\textbf{P}}_r\begin{pmatrix}
    \frac{u_z}{z}\\
    \frac{u_r}{r}\\
    \sigma_{rr}
    \end{pmatrix}
        =0 \nonumber
\end{equation}
at the surface ($r=r_s$), hence
\begin{equation}\label{rrcondition}
    ^t\textsf{\textbf{P}}_r \textsf{\textbf{M}}_s(\rho_s)
    \begin{pmatrix}
    C_s\\
    A_s\\
   B_s
    \end{pmatrix}
        =0
\end{equation}

The last condition, on $\sigma_{zz}$ integrated over the NW cross
section, is
\begin{equation}\label{zzcondition}
   \sum_{i=0}^s \eta_i~^t\textsf{\textbf{P}}_{iz}
    \begin{pmatrix}
    C_i\\
    A_i\\
    B_i
    \end{pmatrix}
       =0
\end{equation}
with $^t\textsf{\textbf{P}}_{iz}=\begin{pmatrix}
    c_{33}&
    2c_{13}&
    0
    \end{pmatrix}$
written with the values of stiffness constants appropriate to the
material in layer $i$. Combining Eq.~\ref{recurrence} to
\ref{zzcondition} we obtain a set of two linear equations for $A_0$
and $C_0$.

\subsubsection {The case of uniform Poisson ratios}

If we assume a common value of the Poisson ratios in the different
materials, as in ref.~\cite{Aifantis} (\emph{i.e.}, if
$c_{ijkl}^i/c_{ijkl}^0$ takes a single value $\chi_i$), several
explicit expressions are obtained.

With this assumption, the continuity of $\sigma_{rr}$ at the
interface at $r_i$ (last line of Eq.~\ref{interface}, multiplied by
$\rho_i^2$ ) is
\begin{eqnarray}\label{}
    \chi_i \left[c_{13}C_i \rho_i^2+ (c_{11}+c_{12})A_i \rho_i^2- (c_{11}-c_{12})B_i\right]=
    \nonumber\\ \chi_{i+1} \left[c_{13}C_{i+1} \rho_i^2+(c_{11}+c_{12})A_{i+1} \rho_i^2-(c_{11}-c_{12})B_{i+1}\right]\nonumber \\
\end{eqnarray}
Adding these equations for all interfaces, including the surface for
which the right-hand member is zero, and using
$\eta_i=(\rho_i^2-\rho_{i-1}^2)/\rho_s^2$, we obtain
\begin{equation}\label{}
    c_{13} \sum_{i=0}^s C_i\chi_i \eta_i+(c_{11}+c_{12}) \sum_{i=0}^s A_i \chi_i
    \eta_i=0 \nonumber
\end{equation}
The second condition is that the integral of $\sigma_{zz}$ over the
NW cross-section vanishes:
\begin{equation}\label{}
    c_{33} \sum_{i=0}^s C_i \chi_i \eta_i+2c_{13} \sum_{i=0}^s A_i \chi_i \eta_i=0 \nonumber
\end{equation}
Hence the two sums must vanish independently
\begin{eqnarray} \label{summation}
  \sum_{i=0}^s C_i \chi_i \eta_i=0 \nonumber\\
  \sum_{i=0}^s A_i \chi_i \eta_i=0
\end{eqnarray}

Another simple result is obtained for the strain along the axis. The
first line of Eq.~\ref{interface} or \ref{recurrence},
$C_{i+1}=C_i-f_{\| i+1}+f_{\| i}$, results in $C_i=C_0-f_{\| i}$ and
finally, using the first sum rule $\sum_{i=0}^s C_i \chi_i
\eta_i=0$,
\begin{equation}\label{}
   C_i= \sum_{j=0}^s \chi_j  \eta_j f_{\| j} - f_{\| i}
\end{equation}

The recurrence on the in-plane strain is not as simple. Indeed the
transfer matrix in Eq.~\ref{recurrence} is
\begin{eqnarray}
  &&\textsf{\textbf{M}}^{-1}_{i+1}(\rho_i) \textsf{\textbf{M}}_i(\rho_i) = \textbf{1}
  \nonumber\\
   &&+ \frac{\chi_i -\chi_{i+1}}{\chi_{i+1}}\frac{1}{2c_{11}}
   \begin{pmatrix}
    0&0&0\\
    c_{13}&(c_{11}+c_{12})&-(c_{11}-c_{12})\\
    -c_{13}\rho^{2}&-(c_{11}+c_{12})\rho^{2}&(c_{11}-c_{12})\rho^{2}
    \end{pmatrix}
   \nonumber
\end{eqnarray}
which shows that if $\chi_{i+1}\neq \chi_i $, only the recurrence on
$C_i$ is simple.

It is worth however to write the result for the simple core-shell
NW. Simplifying the notation, with $\eta=\eta_0$, $\chi=\chi_s$,
$\chi_0=1$, $f=f_s$),
\begin{eqnarray} \label{strainhexa}
\varepsilon_{zz}^c~~~~&=&\frac{(1-\eta )\chi}{\eta+(1-\eta )\chi}f_{\|}  \nonumber \\
\varepsilon_{zz}^s~~~~&=&\frac{-\eta }{\eta+(1-\eta )\chi}f_{\|}  \nonumber \\
\frac{\varepsilon_{\theta \theta}^c-\varepsilon_{rr}^c}{2}&=&0  \nonumber \\
\frac{\varepsilon_{\theta
\theta}^s-\varepsilon_{rr}^s}{2}&=&B_s\frac{r_c^2}{r^2}  \nonumber \\
\frac{\varepsilon_{\theta
\theta}^c+\varepsilon_{rr}^c}{2}&=&\frac{(1-\eta )\chi}{\eta+(1-\eta
)\chi} (f_{\bot}+B_s)  \nonumber \\
\frac{\varepsilon_{\theta
\theta}^s+\varepsilon_{rr}^s}{2}&=&\frac{-\eta }{\eta+(1-\eta )\chi}
(f_{\bot}+B_s)
\end{eqnarray}
where \begin{eqnarray} \label{parahexa}
B_s&=&-\frac{(c_{11}+c_{12})f_{\bot}+c_{13}f_{\|}}{(c_{11}-c_{12})[\eta+(1-\eta
)\chi]+(c_{11}+c_{12})} \nonumber
\\f_{\bot}+B_s&=&\frac{(c_{11}-c_{12})[\eta+(1-\eta
)\chi]f_{\bot}-c_{13}f_{\|}}{(c_{11}-c_{12})[\eta+(1-\eta
)\chi]+(c_{11}+c_{12})}
\end{eqnarray}

\subsubsection {The case of a uniform stiffness tensor}

If all materials have the same values of stiffness constants (all
$\chi_i=1$), all $\textsf{\textbf{M}}_i$ matrices are identical. The
recurrence relation (Eq.~\ref{recurrence}) is simply
\begin{eqnarray}\label{recurrence1}
\begin{pmatrix}
    C_{i+1}\\
    A_{i+1}\\
    B_{i+1}
    \end{pmatrix}
    &=&
 \begin{pmatrix}
    C_i\\
    A_i\\
    B_i
    \end{pmatrix}
    \nonumber\\
    &&-    \textsf{\textbf{M}}^{-1}(\rho_i)
\left[
\begin{pmatrix}
    f_{\|(i+1)}\\
    f_{\bot (i+1)}\\
    0
    \end{pmatrix}
    -
    \begin{pmatrix}
    f_{\|i}\\
    f_{\bot}i\\
    0
    \end{pmatrix} \right]
    \end{eqnarray}
or
\begin{eqnarray} \label{recurrence2}
  C_{i+1} &=& C_i+f_{\|i}-f_{\|(i+1)} \nonumber\\
  A_{i+1} &=& A_i+\frac{c_{11}-c_{12}}{2c_{11}}\left(
f_{\bot i}- f_{\bot (i+1)}\right)
-\frac{c_{13}}{2c_{11}}\left(f_{\|i}-f_{\|(i+1)}\right)\nonumber\\
  B_{i+1} &=& B_i+\frac{c_{11}+c_{12}}{2c_{11}}\left(
f_{\bot i}-f_{\bot (i+1)}\right)\rho_i^2\nonumber\\
  &+&\frac{c_{13}}{2c_{11}}\left(f_{\|i}-f_{\|(i+1)}\right)\rho_i^2
\end{eqnarray}
so that
\begin{eqnarray}
  C_i &=& C_0-f_{\| i} \nonumber\\
  A_i &=& A_0-\frac{c_{11}-c_{12}}{2c_{11}}\
f_{\bot i}
+\frac{c_{13}}{2c_{11}}f_{\| i}  \nonumber\\
   B_i &=&
\frac{c_{11}+c_{12}}{2c_{11}} \left[\sum_{j=0}^{i-1} \eta_j f_{\bot
j} \rho_s^2- f_{\bot i}\rho_{i-1}^2 \right] \nonumber
\\&+&\frac{c_{13}}{2c_{11}}\left[\sum_{j=0}^{i-1} \eta_j f_{\| j}\rho_s^2 -
f_{\| i}\rho_{i-1}^2 \right]\nonumber \\
\end{eqnarray}
and finally, using Eq.~\ref{summation}
\begin{eqnarray}  \label{}
C_i&=& \sum_{j=0}^s \eta_j f_{\| j} - f_{\| i}  \nonumber \\
A_i &=& \frac{(c_{11}-c_{12})}{2c_{11}} \left[\sum_{j=0}^s \eta_j
f_{\bot j} - f_{\bot i} \right]  \nonumber
\\&-&\frac{c_{13}}{2c_{11}}\left[\sum_{j=0}^s \eta_j f_{\| j} -
f_{\| i} \right]
\end{eqnarray}

The strain configuration is thus:
\begin{eqnarray}  \label{multishell}
(\varepsilon_{zz})_i&=& \sum_{j=0}^s \eta_j f_{\| j} - f_{\| i}  \nonumber \\
(\frac{\varepsilon_{\theta \theta} -\varepsilon_{rr}}{2})_i &=&
\frac{c_{11}+c_{12}}{2c_{11}} \left[\sum_{j=0}^{i-1} \eta_j f_{\bot
j} \left(\frac{r_s}{r}\right)^2- f_{\bot
i}\left(\frac{r_{i-1}}{r}\right)^2 \right] \nonumber
\\&+&\frac{c_{13}}{2c_{11}}\left[\sum_{j=0}^{i-1} \eta_j f_{\| j}\left(\frac{r_s}{r}\right)^2 -
f_{\| i}\left(\frac{r_{i-1}}{r}\right)^2 \right]\nonumber \\
(\frac{\varepsilon_{\theta \theta} +\varepsilon_{rr}}{2})_i &=&
\frac{(c_{11}-c_{12})}{2c_{11}} \left[\sum_{j=0}^s \eta_j f_{\bot j}
- f_{\bot i} \right]  \nonumber
\\&-&\frac{c_{13}}{2c_{11}}\left[\sum_{j=0}^s \eta_j f_{\| j} - f_{\| i}
\right]
\end{eqnarray}
It can be applied to a multishell structure such as in
ref.~\cite{Eymery}.

In the case of the simple core-shell NW, we recover the usual
expressions, Eq.~\ref{StrainSameStiffness}:
\begin{eqnarray} \label{strainhexaSameStiffness}
\varepsilon_{zz}^c&=&(1-\eta )f_{\|}  \nonumber \\
\varepsilon_{zz}^s&=&-\eta f_{\|}
 \nonumber\\
\frac{\varepsilon_{\theta \theta}^s -\varepsilon_{rr}^s}{2} &=& B_s
\frac{r_c^2}{r^2}\nonumber \\
\varepsilon_{\theta \theta}^c =\varepsilon_{rr}^c&=&(1-\eta )
(f_{\bot}+B_s)
 \nonumber\\
\frac{\varepsilon_{\theta \theta}^s +\varepsilon_{rr}^s}{2} &=&-
\eta (f_{\bot}+B_s)
\end{eqnarray}
where
\begin{eqnarray} \label{parahexaSameStiffness}
B_s&=&- \frac{(c_{11}+c_{12})f_{\bot}+c_{13}f_{\|}}{2c_{11}}\nonumber \\
f_{\bot}+B_s&=&\frac{(c_{11}-c_{12})f_{\bot}-c_{13}f_{\|}}{2c_{11}}  \nonumber \\
\end{eqnarray}
A comparison between the expressions for $\varepsilon_{zz}$ in
Eq.~\ref{strainhexa}-\ref{parahexa} and
Eq.~\ref{strainhexaSameStiffness}-\ref{parahexaSameStiffness}
illustrates the effect of a different hardness of the two materials:
in the sharing of lattice mismatch, the weight is defined by the
area ratio multiplied by the hardness ratio. In particular, for a
thin layer ($\eta\approx1$), the strain in the core is multiplied by
$\chi$, while for a thick layer ($\eta \ll 1$), the strain in the
shell is divided by $\chi$.

Note also that Eq.~\ref{multishell} can be used to describe a
continuous distribution in a NW, just by replacing the discrete sums
by integrals:
\begin{eqnarray}  \label{}
C_0&=& \sum_{j=0}^s \eta_j f_{\| j} \rightarrow \int  f_{\| } (r)\frac{2 r dr}{r_s^2} \nonumber \\
A_0 &=& \frac{(c_{11}-c_{12})}{2c_{11}} \sum_{j=0}^s \eta_j f_{\bot
j}  -\frac{c_{13}}{2c_{11}} \sum_{j=0}^s \eta_j f_{\| j}\nonumber \\
&\rightarrow & \frac{(c_{11}-c_{12})}{2c_{11}} \int f_{\bot}
(r)\frac{2 r dr}{r_s^2}
-\frac{c_{13}}{2c_{11}} \int f_{\| } (r)\frac{2 r dr}{r_s^2}\nonumber \\
\end{eqnarray}
An interesting consequence is that the strain in the core of a
core-shell NW (for instance, GaAs-Ga$_{1-x}$Al$_x$As) or in a
multishell NW, is determined by the  composition integrated over the
shell(s), and not by the exact distribution within the shell(s). The
analogy with the Gauss theorem of electrostatics is not fortuitous
and it has been discussed in Ref.~\cite{Davies} for a quantum dot
buried in an isotropic material. However it applies only in special
cases where the (vectorial) Lam\'{e} - Clapeyron - Navier equation
can be mapped onto the scalar Poisson equation, with the local
mismatch defining the equivalent of the electric charge.
\subsubsection{Summary and electronic properties.}
To sum up, the strain configuration in a core-shell NW grown along
the hexagonal direction of wurtzite crystal is transversely
isotropic. It is given by
Eq.~\ref{strainhexaSameStiffness}-\ref{parahexaSameStiffness} if the
two materials have the same hardness, and
Eq.~\ref{strainhexa}-\ref{parahexa} for a hardness ratio $\chi\neq
1$. An explicit expression,  Eq.~\ref{multishell}, also exists for a
multishell NW if the stiffness constants are identical over the NW.

The potential configuration for the bottom of the conduction band
and the top of the valence band near the center of the Brillouin
zone is obtained from these expressions using the Bir-Pikus
phenomenological coupling \cite{deformation}. In the core, the
non-vanishing strain components are $\varepsilon_{zz}$ and
$\frac{1}{2}(\varepsilon_{rr}+\varepsilon_{\theta \theta})$ so that
the Bir-Pikus Hamiltonian \cite{deformation} has only diagonal
elements in the usual basis quantized along the $c$-axis. Note
however that the resulting matrix elements may be of the same order
as the other terms describing the top of the valence band and the
excitons (spin-orbit coupling, crystal field splitting and exchange
terms). In the shell (s), the in-plane shear strain
$\frac{1}{2}(\varepsilon_{rr}-\varepsilon_{\theta \theta})$
introduces non-diagonal terms, which mix the valence band states
initially quantized along the $c$-axis. Actually, this term may give
the main contribution to the hole potential in the shell.
Interestingly, it splits the hole multiplet in such a way that one
type of holes is confined in the vicinity of the interface, far from
the sidewall.

The piezoelectric effect is described by an axial polarization,
determined by the two strain components $\varepsilon_{zz}$ and
$\frac{1}{2}(\varepsilon_{rr}+\varepsilon_{\theta \theta})$. There
is no coupling to the in-plane shear strain.

The present study also confirms that GaN-InN multi-quantum-well NWs
\cite{Eymery} should indeed feature no built-in piezoelectric field
perpendicular to the QWs, but an in-plane shear-strain different
from well to well.
\subsection {Application to real systems}
\subsubsection{GaN-AlN nanowires}
The strain in the core of single GaN-AlN core-shell NWs grown by
plasma-assisted molecular beam epitaxy was measured by resonant
x-ray diffraction, Raman spectroscopy and high resolution
transmission electron microscopy \cite{Herstroffer}: for unrelaxed
NWs, it favorably compares to the results of a microscopic
calculation using the valence-force-field model, and to a
macroscopic calculation assuming uniform strain along the $c$-axis
and vanishing strain in the plane. Complementary results are given
in Ref.~\cite{Rigutti}.

\begin{figure}
\resizebox{0.5\textwidth}{!}{\includegraphics {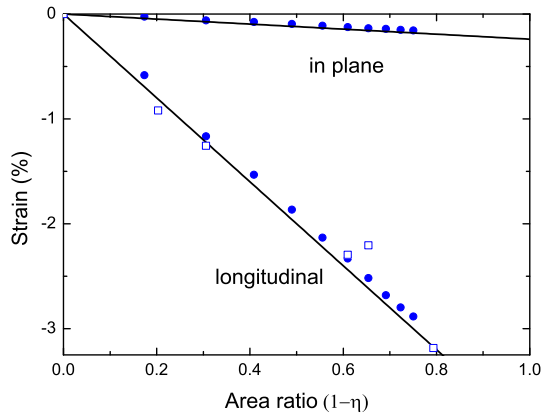}}
\caption{Strain in the core of a GaN-AlN NW, as a function of the
area ratio. Solid symbols, microscopic calculation, open symbols,
electron microscopy data, both from ref.~\cite{Herstroffer}; lines,
present calculation.\label{GaN}}
\end{figure}

The stiffness constants of GaN and AlN \cite{Polian,Wright,McNeil}
are quite similar, hence we take $\chi=1$. The lattice mismatch is
slightly anisotropic, $f_{\bot}=-2.5\%$ and $f_{\|}=-4.0\%$. The
present calculation predicts a uniform strain in the core,
$\varepsilon_{zz}^c=f_{\bot}(1-\eta)$ along the NW axis and
$\frac{1}{2}(\varepsilon_{\theta
\theta}^c+\varepsilon_{rr}^c)=-(1-\eta)\times 0.25\%$ in the plane.
The agreement with the results of ref.~\cite{Herstroffer} is
excellent, see fig.~\ref{GaN}.

Note that the small value of the in-plane strain is due to a
compensation between the Poisson effect of the longitudinal mismatch
$f_{\|}$ and the direct effect of the in-plane mismatch $f_{\bot}$,
see $f_{\bot}+B_s$ in Eq.~\ref{parahexaSameStiffness}.

\subsubsection{ZnO nanowires}
\begin{figure}
\resizebox{0.5\textwidth}{!}{\includegraphics {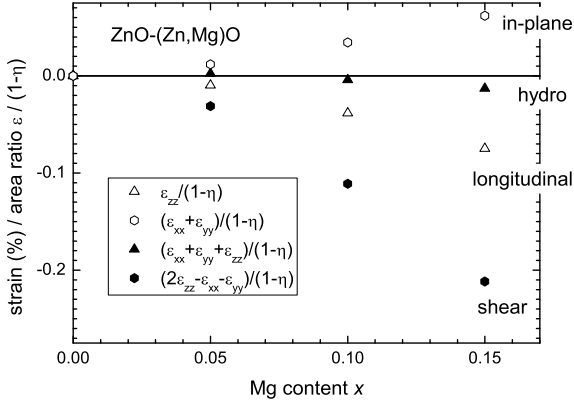}}
\caption{Strain in the core of a ZnO-(Zn,Mg)O NW, divided by the
shell/NW area ratio $(1-\eta)$, as a function of the Mg content $x$.
The values of the lattice mismatch along the NW axis and
perpendicular to it are the experimental values of
ref.~\cite{Herstroffer} and are opposite in sign. Open symbols,
longitudinal and in-plane strain. Full symbols, shear strain and
volume change.\label{ZnO}}
\end{figure}

ZnO is such that $c_{11}-c_{12}<c_{13}$ \cite{Azuhata}: for an
isotropic lattice mismatch, the Poisson effect prevails in the
in-plane strain. ZnO cores are  often associated to a strongly
mismatched shell and in this case the structure is no more coherent.
A moderate mismatch exists in ZnO-(Zn,Mg)O. According to a
synchrotron x-ray study of polycrystalline wurtzite (Zn,Mg)O
\cite{Kim}, it is strongly anisotropic, with $f_{\bot}$ and $f_{\|}$
opposite in sign: this is attributed to a change in the ionicity. As
a result (fig.~\ref{ZnO}), the core experiences a significant shear
strain with a ten times smaller volume change. In other words, the
$c/a$ ratio, which represents the deviation from "ideal" wurtzite,
is changed at almost constant volume. Note that a non linear
character of the piezoelectric effect has been measured in CdTe
\cite{Andre} and predicted for other semiconductors as well
\cite{Bester}. As it is attributed to a dependence of the
piezoelectric coefficient on the hydrostatic strain, this non-linear
character should not show up in a ZnO-(Zn,Mg)O NW.

\section{Cubic semiconductors along $<111>$}
We now consider NWs of semiconductors with the zinc-blende or
diamond structure, grown along a trigonal axis. The (111) plane is
known to be isotropic with respect to some mechanical properties, so
that the cylindrical approximation is quite natural for such NWs. We
use it first, and compare its results to data known for real
systems. However the shear strain present in the shell gives rise to
warping, with a 3-fold symmetry, which is calculated analytically in
section \ref{warping}.
\subsection{Cylindrical approximation}
\subsubsection{Calculation}
If the parameter $c$ is not zero, the stiffness tensor must be
calculated in the relevant axes. It can be done on the $c_{ijkl}$
tensor, or directly in the Voigt notation using the rotation rules
described in ref.~\cite{Wortman}. We take the basis defined by the
three vectors $x=[1\bar{1}0]$, y=$[11\bar{2}]$, z=$[111]$, identical
to that in ref.~\cite{Gronqvist} but different from
ref.~\cite{Boxberg}. Then \cite{Nye} the stiffness matrix is
\begin{equation}\label{stiffnessTensor111}
    \begin{pmatrix}
\tilde{c}_{11}&\tilde{c}_{12}&\tilde{c}_{13}&\tilde{c}_{14}&0&0 \\
\tilde{c}_{12}&\tilde{c}_{11}&\tilde{c}_{13}&-\tilde{c}_{14}&0&0 \\
\tilde{c}_{13}&\tilde{c}_{13}&\tilde{c}_{33}&0&0&0 \\
\tilde{c}_{14}&-\tilde{c}_{14}&0&\tilde{c}_{44}&0&0 \\
0&0&0&0&\tilde{c}_{44}&\tilde{c}_{14} \\
0&0&0&0&\tilde{c}_{14}&\frac{\tilde{c}_{11}-\tilde{c}_{12}}{2} \\
\end{pmatrix}
\end{equation}
The six components are not independent since they can be expressed
using the three coefficients $c_{11}$, $c_{12}$ and $c_{44}$
relevant for the cubic symmetry \cite{Gronqvist}:
\begin{eqnarray} \label{stiffness111}
    \tilde{c}_{11}&=c_{11}-\frac{1}{2}c=\frac{1}{2}c_{11}+\frac{1}{2}c_{12}+c_{44} \nonumber \\
    \tilde{c}_{33}&=c_{11}-\frac{2}{3}c=\frac{1}{3}c_{11}+\frac{2}{3}c_{12}+\frac{4}{3}c_{44} \nonumber \\
    \tilde{c}_{12}&=c_{12}+\frac{1}{6}c=\frac{1}{6}c_{11}+\frac{5}{6}c_{12}-\frac{2}{6}c_{44} \nonumber \\
    \tilde{c}_{13}&=c_{12}+\frac{1}{3}c=\frac{1}{3}c_{11}+\frac{2}{3}c_{12}-\frac{2}{3}c_{44} \nonumber \\
    \tilde{c}_{44}&=c_{44}+\frac{1}{3}c=\frac{1}{3}c_{11}-\frac{1}{3}c_{12}+\frac{1}{3}c_{44} \nonumber \\
    \tilde{c}_{14}&=\frac{1}{3\sqrt{2}}c=\frac{1}{3\sqrt{2}}(c_{11}-c_{12}-2c_{44})
\end{eqnarray}
The stiffness tensor reflects the threefold symmetry of the trigonal
axis: it is quite similar to that of the wurtzite structure along
the $c$-axis. However there is a set of additional terms,
$\tilde{c}_{14}$. To better understand these terms, we can write the
stiffness matrix in the $\textbf{e}_r$, $\textbf{e}_{\theta}$,
$\textbf{e}_z$ axes, rotated with respect to the previous one by an
angle $\theta$ around the $<111>$ (or $\textbf{e}_z$) axis:
\begin{eqnarray}\label{stiffnessTensor111Rotated}
&&\begin{pmatrix}
\tilde{c}_{11}&\tilde{c}_{12}&\tilde{c}_{13}&0&0&0 \\
\tilde{c}_{12}&\tilde{c}_{11}&\tilde{c}_{13}&0&0&0 \\
\tilde{c}_{13}&\tilde{c}_{13}&\tilde{c}_{33}&0&0&0 \\
0&0&0&\tilde{c}_{44}&0&0 \\
0&0&0&0&\tilde{c}_{44}&0 \\
0&0&0&0&0&\frac{\tilde{c}_{11}-\tilde{c}_{12}}{2} \\
\end{pmatrix} +\nonumber\\
&& \tilde{c}_{14}
\begin{pmatrix}
0&0&0&\cos3\theta&\sin3\theta&0 \\
0&0&0&-\cos3\theta&-\sin3\theta&0 \\
0&0&0&0&0&0 \\
\cos3\theta&-\cos3\theta&0&0&0&\sin3\theta \\
\sin3\theta&-\sin3\theta&0&0&0&\cos3\theta \\
0&0&0&\sin3\theta&\cos3\theta&0 \\
\end{pmatrix}
\end{eqnarray}
The trigonal symmetry of the $\tilde{c}_{14}$ terms is clear, as
noted in Ref~\cite{Schulz}. Note that these contributions average to
zero over a complete $2\pi$-turn. Moreover, they are quite small:
for instance in GaAs,
$\tilde{c}_{14}/(\tilde{c}_{11}+2\tilde{c}_{12})=-0.05$.

The complete Lam\'{e} - Clapeyron - Navier equation in the $x$, $y$,
$z$ basis, Eq.~\ref{LCNgeneral}, is written in Appendix
\ref{completeLCN}. Omitting terms excluded by the invariance by
translation, we obtain:
%\begin{widetext}
\begin{eqnarray} \label{LCN111}
    &&\frac{\tilde{c}_{11}-\tilde{c}_{12}}{2} (\frac{\partial^2}{\partial
y^2}+\frac{\partial^2}{\partial x^2})u_x+
\frac{\tilde{c}_{11}+\tilde{c}_{12}}{2}
\frac{\partial}{\partial x}(\frac{\partial u_x}{\partial x}+\frac{\partial u_y}{\partial y})\nonumber \\
    &&~~~~+2\tilde{c}_{14}(\frac{\partial ^2u_x}{\partial y \partial z}+ \frac{\partial ^2u_y}{\partial z \partial x}+ \frac{\partial ^2u_z}{\partial x \partial y})=0\nonumber \\
    &&\frac{\tilde{c}_{11}-\tilde{c}_{12}}{2} (\frac{\partial^2}{\partial
y^2}+\frac{\partial^2}{\partial x^2})u_y+
\frac{\tilde{c}_{11}+\tilde{c}_{12}}{2}
\frac{\partial}{\partial y}(\frac{\partial u_x}{\partial x}+\frac{\partial u_y}{\partial y})\nonumber\\
    &&~~~~+\tilde{c}_{14}(2\frac{\partial ^2u_x}{\partial z \partial x}+\frac{\partial ^2u_z}{\partial x^2}-2\frac{\partial ^2u_y}{\partial y \partial z}-\frac{\partial ^2u_z}{\partial y^2})=0\nonumber\\
    &&\tilde{c}_{44}(\frac{\partial^2}{\partial y^2}+\frac{\partial^2}{\partial
x^2})u_z+\tilde{c}_{14}(2\frac{\partial ^2u_x}{\partial x \partial y}+ \frac{\partial ^2u_y}{\partial x^2}- \frac{\partial ^2u_y}{\partial y^2})=0\nonumber\\
\end{eqnarray}
%\end{widetext}
The effect of the $\tilde{c}_{14}$ terms will be described in
Section \ref{warping}. Ignoring these terms for a while, the
equation is the same as in the wurtzite case. Then the solution is
obtained by replacing the $c_{ij}$ in
Eq.~\ref{strainhexa}-\ref{parahexa} by the $\tilde{c}_{ij}$ and
their expression (Eq.~\ref{stiffness111}). The result is identical
to Eq.~\ref{strainhexa}, but with $f_{\|}=f_{\bot}=f$, and
\begin{eqnarray}\label{para111}
&&B_s=\nonumber\\
&&- \frac{3(c_{11}+2c_{12})}{(c_{11}-c_{12}+4c_{44})[\eta+(1-\eta
)\chi]+(2c_{11}+4c_{12}+2c_{44})}f\nonumber\\
&&f+B_s=\nonumber\\
&&\frac{(c_{11}-c_{12}+4c_{44})[\eta+(1-\eta
)\chi]-(c_{11}+2c_{12}-2c_{44})}{(c_{11}-c_{12}+4c_{44})[\eta+(1-\eta
)\chi]+(2c_{11}+4c_{12}+2c_{44})}f \nonumber\\
\end{eqnarray}

If the stiffness constants are identical in the two materials, we
recover the same expression as above
(Eq.~\ref{StrainSameStiffness}), with :
\begin{eqnarray} \label{strain111identical}
B_s&=&-f \frac{c_{11}+2c_{12}}{c_{11}+c_{12}+2c_{44}} \nonumber \\
(f+B_s)&=& f \frac{-c_{12}+2c_{44}}{c_{11}+c_{12}+2c_{44}}
\end{eqnarray}

In the core, the strain corresponds to a uniform hydrostatic strain
$\varepsilon_{\texttt{hydro}}=\varepsilon_{zz}+\varepsilon_{rr}+\varepsilon_{\theta\theta}$
and a uniform trigonal shear strain
$\varepsilon_{\texttt{shear}}=2\varepsilon_{zz}-\varepsilon_{rr}-\varepsilon_{\theta\theta}$.
It should be kept in mind that the axis used in these expression are
$x=[1\bar{1}0]$, $y=[11\bar{2}]$, $z=[111]$; in the cubic axes,
$x'$, $y'$, $z'$, the previous results means, for the core,
$\varepsilon_{x'x'}=\varepsilon_{y'y'}=\varepsilon_{z'z'}=\frac{1}{3}\varepsilon_{\texttt{hydro}}$
and
$\varepsilon_{x'y'}=\varepsilon_{y'z'}=\varepsilon_{z'x'}=\frac{1}{6}\varepsilon_{\texttt{shear}}$.

\subsubsection{Excitons} \label{exciton}
Finally, we consider the exciton energy in the core of a core-shell
NW, in the absence of confinement effects. In a strained
semiconductor, it is expected at
$E_X=E_X^0-(a'+a)~\varepsilon_{\texttt{hydro}}\pm \frac{1}{2} b
~\varepsilon_{\texttt{shear}}$ (for a tetragonal shear strain) or
$E_X=E_X^0-(a'+a)~ \varepsilon_{\texttt{hydro}}\pm
\frac{1}{2}(d/\sqrt{3})~\varepsilon_{\texttt{shear}}$ (for a
trigonal shear strain). The coefficient $a'$ describes the coupling
of conduction electrons to strain, and $a$, $b$, $d$ describe the
coupling of holes (Bir-Pikus Hamiltonian \cite{deformation}). Using
Eq.~\ref{strain111identical} we obtain
\begin{eqnarray}\label{exciton111}
    E_X=E_X^0-(a'+a) \frac{(c_{11}-c_{12}+6c_{44})}{c_{11}+c_{12}+2c_{44}}(1-\eta)f \nonumber \\ \pm
\frac{d}{\sqrt{3}}\frac{(c_{11}+2c_{12})}{c_{11}+c_{12}+2c_{44}}(1-\eta)f
\end{eqnarray}
The sign $+$ is for the exciton formed with the light hole (moment
$\pm\frac{1}{2}$ along the NW axis), the sign - for the heavy hole
($\pm\frac{3}{2}$) exciton. A more complete analysis is given at the
end of section \ref{warping}.
\subsection{Application to real systems}
\subsubsection{GaAs-based nanowires}
\begin{figure}
\resizebox{0.5\textwidth}{!}{\includegraphics {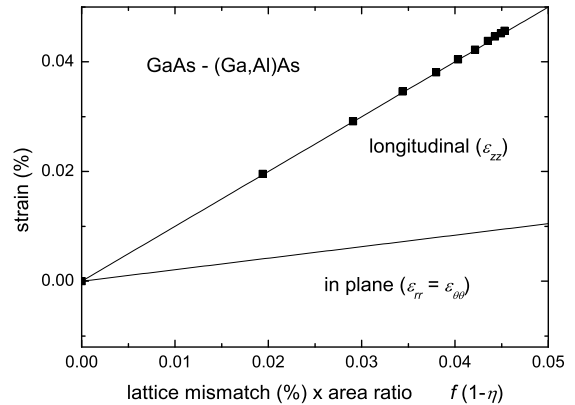}}
\caption{Strain in a GaAs core, as a function of area ratio
$(1-\eta)$ times the lattice mismatch $f$. Symbols are the numerical
calculation of ref.~\cite{Hocevar}, lines are the present
calculation. \label{GaAs}}
\end{figure}

The calculation for a GaAs-Ga$_{0.65}$Al$_{0.35}$As NW with
hexagonal cross section, using the valence force field model,
fig.~3c of ref.~\cite{Hocevar}, fully agrees (fig.~\ref{GaAs}) with
the present value $\varepsilon_{zz}^c=(1-\eta )f$. The in-plane
strain is "four times smaller" \cite{Hocevar}, which also agrees
with the ratio 0.22 obtained in the present calculation  using the
stiffness constants of (Ga,Al)As \cite{Adachi}, with $\chi=1$.

\begin{figure}
\resizebox{0.5\textwidth}{!}{\includegraphics {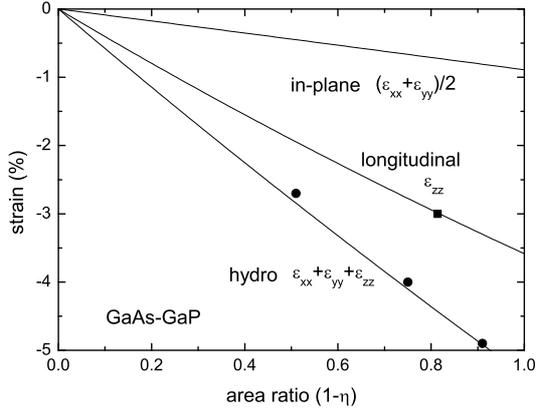}}
\caption{Strain in a GaAs core with a GaP shell, as a function of
area ratio $(1-\eta)$. Symbols are the numerical calculation of
ref.~\cite{Gronqvist} (square) and ref.~\cite{Montazeri} (circles),
lines are the present calculation. \label{GaAsGaP}}
\end{figure}

In GaAs-GaP NWs, the lattice mismatch is 3.6\%, and the stiffness
constants differ by a factor $\chi\approx 1.10$ to 1.17
\cite{Vurgaftman}. NWs with either a circular or a hexagonal cross
sections have been modeled by Gr\"{o}nqvist \emph{et al.}
\cite{Gronqvist} using both the valence force field model and a
finite element treatment of the continuum elasticity theory. Other
core-shell configurations with hexagonal cross-sections are
described in ref.~\cite{Montazeri}. In the case of a circular
cross-section, it confirms the present result that the axial strain
is uniform in the core and in the shell, and that the in-plane
strain is also uniform in the core. Using the appropriate values of
the area ratio, and the stiffness constant values of GaP
\cite{Vurgaftman} with an average ratio $\chi=1.14$ for GaAs-GaP),
we calculate the solid lines shown in fig.~\ref{GaAsGaP}, in good
agreement with numerical calculations. Note the small but visible
bowing which is due to the different values of the stiffness
constants in GaAs and GaP.
\subsubsection{InAs-based nanowires}
\begin{figure}
\resizebox{0.5\textwidth}{!}{\includegraphics  {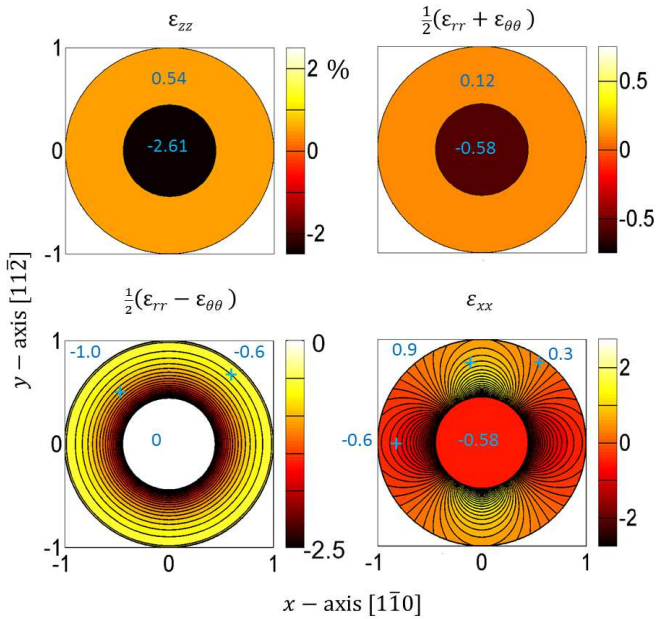}}
 \caption{Strain maps for an InAs core with
an InP shell, with the same area ratio $\eta=0.2$ as in
ref.~\cite{Boxberg}. The lattice mismatch is $f=-3.15\%$, stiffness
ratio $\chi=1.2$, and $B_s\tilde{c}_{14}/\tilde{c}_{44}=-0.82\%$.
Contour line spacings are $0.1\%$. The shear component
$\varepsilon_{zz}$ in (a) and
$\frac{1}{2}(\varepsilon_{xx}+\varepsilon_{yy})=\frac{1}{2}(\varepsilon_{rr}+\varepsilon_{\theta
\theta})$ in (b) are uniform in the core and in the shell. A shear
strain $\frac{1}{2}(\varepsilon_{rr}-\varepsilon_{\theta \theta})$
is present in the shell; it is rotationally invariant (c). Plotting
$\varepsilon_{xx}$ introduces an apparent dependence on $\theta$.
Maps of $\varepsilon_{xx}-\varepsilon_{yy}$, $2\varepsilon_{xy}$ and
$\varepsilon_{yy}-\varepsilon_{xx}$ are identical, but for a
rotation by $\pi/4$ or $\pi/2$. Maps (a), (b) and (d) can be
compared to ref.~\cite{Boxberg}. \label{isoStrainMap}}
\end{figure}

Similar results are obtained in InAs-InP NWs. They favorably compare
(fig.~\ref{isoStrainMap}) with the results of numerical calculations
\cite{Boxberg}. We will come back to this system in the section on
warping (\ref{warping}).

In GaAs-InAs NWs, the approximation of a constant Poisson ratio is
not reasonable and a direct inversion of the full matrix
(Eq.~\ref{fullHexa} where the $c_{ij}$ have been replaced by the
$\tilde{c}_{ij}$, Eq.~\ref{stiffness111}), or the equivalent
transfer matrix method, should be used.
\subsubsection{ZnTe nanowires}
Photoluminescence and cathodoluminescence have been measured on
ZnTe-(Zn,Mg)Te core-shell NWs \cite{Wojn12}, with a peak at 2.31eV,
\emph{i.e.}, a 60 meV redshift with respect to the exciton in bulk
ZnTe; this is a large shift, larger than usually observed in
strained 2D layers. In bare ZnTe NWs \cite{Artioli}, a small (3 meV)
blueshift is observed.

\begin{figure}
\resizebox{0.5\textwidth}{!}{\includegraphics {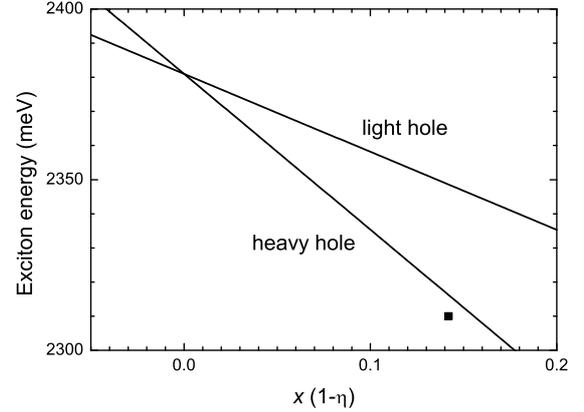}}
\caption{Exciton energy in a ZnTe NW with a (Zn,Mg)Te shell, as a
function of area ratio $(1-\eta)$ times the Mg content $x$.
Confinement effects are ignored. \label{ZnTe}}
\end{figure}

The values of the deformation potentials in ZnTe are
\cite{Wardzynski,LeSi89} $a$=5.3~eV and $d/\sqrt3$=2.5~eV, and the
values of the stiffness constants \cite{Berlincourt},
$c_{11}$=73.7~GPa, $c_{12}$=42.3~GPa, and $c_{44}$=32.1~GPa. Then
the excitonic emission of a $\langle 111 \rangle$ oriented cubic
ZnTe NW is (in meV, with $f$ in \%) $E_{NW}=2381-88 (1-\eta)f$ for
the heavy hole and $E_{NW}=2381-44 (1-\eta)f$ for the light hole.
Note the large shift of the heavy-hole exciton, in sharp contrast
with the case of a 2D epitaxial where the effect of the hydrostatic
strain and the shear strain almost compensate. The heavy hole is the
ground state, as found experimentally in ref.~\cite{Wojn12}. For the
NWs studied in ref.~\cite{Wojn12}, with $r_c=$~35~nm, $r_s=$~65~nm,
and $f=1.04\%$ corresponding to the lattice mismatch between a ZnTe
core and a Zn$_{0.8}$Mg$_{0.2}$Te shell \cite{Hartmann}, we obtain
2.31~eV for the heavy-hole exciton, in agreement with the observed
PL line, see fig.~\ref{ZnTe}. The small blueshift observed in bare
ZnTe NW was attributed to a small residual strain due to a thin
oxide shell \cite{Artioli}.

\subsection{Deviations from cylindrical symmetry}
In this section we discuss the two simplifying assumptions which
allow us to derive the previous analytical expressions: (1) NWs have
a circular cross section, and (2) in the NWs with the zinc-blende
structure, the deviation from cylindrical symmetry is small.
\subsubsection{Facets}
Most of the numerical calculations consider NWs with an hexagonal
cross-sections, and actual NWs exhibit more or less well-defined
facets. The present calculation does not reproduce the inhomogeneity
of the in-plane strain which is calculated for a hexagonal NW, but
it was already noted \cite{Pistol} that the central values of strain
are quite similar in hexagonal and circular cores. This was
confirmed in the very detailed study of ref.~\cite{Gronqvist}, where
NWs with hexagonal and circular cross sections are compared. Indeed
the results of the present model compare fairly well to the results
of numerical calculations made for hexagonal NWs. Other approaches
are reviewed in Ref.~\cite{Maranganti} for the case of nanowires
embedded in an infinite or semi-infinite material.
\subsubsection{Warping terms}\label{warping}
The cylindrical symmetry is exact in the case of NWs with the
wurtzite structure, with the $c$-axis along the NW. It is not for
NWs with the zinc-blende structure. As a result, the shell is
warped, as evidenced in the numerical treatment of
ref.~\cite{Gronqvist}. We now describe the analytical calculation of
this additional contribution.

\begin{figure}
\resizebox{0.4\textwidth}{!}{\includegraphics {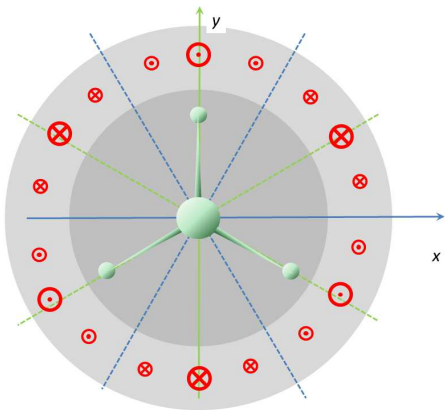}}
\caption{Warping body forces in the shell of a core-shell NW with
$f<0$, grown along $<111>$, due to the trigonal symmetry of the
zinc-blende structure. The shell is pushed upward and downward
according to the red arrows. The tetrahedron of the atomic structure
is shown in green.\label{excessStrain}}
\end{figure}
\begin{figure}
\resizebox{0.5\textwidth}{!}{\includegraphics {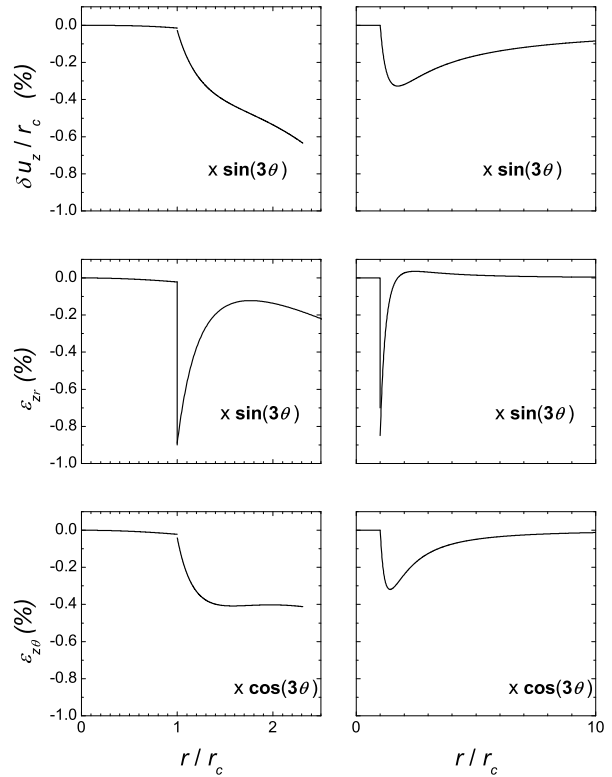}}
\caption{Warping strain due to the trigonal symmetry of the
zinc-blende structure in an InAs-InP core shell NW. The lattice
mismatch is $f=-3.15\%$, stiffness ratio $\chi=1.2$, and $B_s
\tilde{c}_{14}/\tilde{c}_{44}=-0.82\%$. The radial dependence of the
displacement along the axis (top), the radial-axial shear strain
(middle) and the tangential-axial shear strain (bottom) are shown
for a NW with finite shell radius (left column)and infinite shell
radius (right column). The trigonal symmetry appears through the
$\sin 3\theta$ or $\cos 3\theta $ factor, as indicated.
\label{axialShear}}
\end{figure}
Indeed, when calculating the stress corresponding to the cylindrical
strain configuration, additional components appear through the
$\tilde{c}_{14}$ terms in the stiffness tensor: for instance, at the
interface in the $y$-direction ($x=0$, $y=r_c$), a stress component
normal to the interface and surface,
$\sigma_{yz}=\tilde{c}_{14}(\varepsilon_{xx}-\varepsilon_{yy})+2\tilde{c}_{44}\varepsilon_{zy}$,
takes a finite value if we use the strain of Eq.~\ref{strainhexa}
and \ref{para111}, or \ref{strainhexaSameStiffness} and
\ref{strain111identical}. We thus expect an additional strain to
appear,
$\varepsilon_{zy}=-\tilde{c}_{14}(\varepsilon_{xx}-\varepsilon_{yy})/2\tilde{c}_{44}$
where $(\varepsilon_{xx}-\varepsilon_{yy})$ is taken from
Eq.~\ref{strainhexa} and \ref{para111} or
\ref{strainhexaSameStiffness} and \ref{strain111identical}: it
vanishes in the core (where $B_c=0$), but not in the shell where a
non-uniform shear strain
$(\varepsilon^s_{\theta\theta}-\varepsilon^s_{rr}$) exists. With
$c<0$, and $f<0$ (case of GaAs-GaP, InAs-GaAs, CdTe-ZnTe core-shell
NWs, not ZnTe-(Zn,Mg)Te), we expect a positive $\varepsilon_{zy}$,
\emph{i.e.}, the shell is pushed upward, towards $[111]$. Note that
other non-vanishing stress components are obtained by re-introducing
these warping terms into the calculation of the stress, so that they
are of second order in $\tilde{c}_{14}$.

Using the rotated stifffness tensor,
Eq.~\ref{stiffnessTensor111Rotated}, and forcing the stress
component $\sigma_{rz}$ to be zero (and neglecting a contribution of
second order in $\tilde{c}_{14}$), we obtain
$\varepsilon_{zr}=(\tilde{c}_{14}/2\tilde{c}_{44})(\varepsilon_{\theta
\theta}-\varepsilon_{rr}) \sin3\theta $: the shell is alternately
pushed upward and downward, with the expected trigonal symmetry
(fig.~\ref{excessStrain}).

To calculate the complete strain distribution, we must re-calculate
the displacement field thanks to the Lam\'{e} - Clapeyron - Navier
equation.

When introducing the cylindrical solution $u_r(r)=Ar+B r_c^2/r$,
$u_{\theta}=0$, $u_z(z)$) into the complete equation,
Eq.~\ref{LCN111}, the $\tilde{c}_{14}$ terms vanish everywhere but
in the third equation for the shell. There, $2\partial
^2u_x/\partial x
\partial y+\partial ^2u_y/\partial x^2-\partial
^2u_y/\partial y^2=8 B_s r_c^2 \sin3\theta /r^3$. In a treatment to
first order in $\tilde{c}_{14}$ (\emph{i.e}, in the cubic anisotropy
$c$), we have to find an additional displacement $\delta \textbf{u}$
which is solution of the Lam\'{e} - Clapeyron - Navier equation for
the transversely isotropic NW, with no lattice mismatch (they are
already compensated) but with body forces $F_x=0$, $F_y=0$,
$F_z=\tilde{c}_{14}8 B_s r_c^2 \sin3\theta /r^3$ in the shell.

Thus, $\delta \textbf{u}$ is the \emph{response of an isotropic
system} to an \emph{axial shear strain} \cite{Tsukrov} \emph{of
trigonal symmetry} ($\sim \sin 3\theta$).

The solution is $\delta u_x=0$, $\delta u_y=0$ and $\delta u_z$ such
that
\begin{eqnarray}\label{trigoStrain}
    &&\tilde{c}_{44}(\frac{\partial ^2}{\partial x^2}+\frac{\partial ^2}{\partial
    y^2})\delta u_z=0\nonumber\\
    &&\tilde{c}_{44}(\frac{\partial ^2}{\partial x^2}+\frac{\partial ^2}{\partial
    y^2})\delta u_z+\tilde{c}_{14} 8 B_s r_c^2
    \frac{\sin3\theta }{r^3}=0
\end{eqnarray}
in the core and in the shell, respectively. The result is of
trigonal symmetry and can be written, respectively (see Appendix
\ref{generalShear} for details):
\begin{eqnarray}\label{}
&&\frac{\delta u_z}{r_c}=\frac{\tilde{c}_{14}}{\tilde{c}_{44}} B_s
 \alpha_3^c \frac{r^3}{r_c^3}\sin3\theta  \nonumber\\
&&\frac{\delta u_z}{r_c}=\frac{\tilde{c}_{14}}{\tilde{c}_{44}} B_s
(\frac{r_c}{r}+\alpha_3^s \frac{r^3}{r_c^3}+\alpha_{-3}^s \frac{
r_c^3}{r^3})\sin3\theta  \nonumber\\
\end{eqnarray}
where we have used $\alpha_{-3}^c=0$ (no diverging term), and the
three parameters $\alpha_3^c$, $\alpha_3^s$, $\alpha_{-3}^s$ are
determined by the boundary conditions: the non-trivial boundary
conditions are that $\delta u_z$ and $\sigma_{rz}=\tilde{c}_{14}
\sin 3\theta  (\varepsilon_{rr}-\varepsilon_{\theta \theta
})+\tilde{c}_{44}\frac{1}{2} \partial (\delta u_z) / \partial r$ are
continuous at the interface, and $\sigma_{rz}$ vanishes at the
surface. The final result is:
\begin{eqnarray}\label{trigonalStrain}
\left[ \frac{\delta u_z}{r_c}\right]_c &=& \left [\eta^2 (1-\eta) \frac{r^3}{r_c^3}\right]~\frac{\tilde{c}_{14}}{\tilde{c}_{44}} B_s~\sin3\theta \nonumber\\
\left[ \varepsilon_{rz} \right]_c&=& \frac{3}{2}~\left[\eta^2 (1-\eta) \frac{r^2}{r_c^2} \right] ~\frac{\tilde{c}_{14}}{\tilde{c}_{44}} B_s~\sin3\theta \nonumber\\
\left[ \varepsilon_{\theta z} \right]_c&=&\frac{3}{2}~\left[\eta^2(1-\eta) \frac{r^2}{r_c^2} \right] \frac{\tilde{c}_{14}}{\tilde{c}_{44}} B_s\cos3\theta \nonumber\\
\left[ \frac{\delta u_z}{r_c}\right]_s &=& \left [\frac{r_c}{r}-\frac{r_c^3}{r^3}+\eta^2 (1-\eta) \frac{r^3}{r_c^3}\right]~\frac{\tilde{c}_{14}}{\tilde{c}_{44}} B_s~\sin3\theta \nonumber\\
\left[ \varepsilon_{rz} \right]_s&=& \frac{3}{2}~\left[-\frac{r_c^2}{3r^2}+\frac{r_c^4}{r^4}+\eta^2 (1-\eta) \frac{r^2}{r_c^2} \right] ~\frac{\tilde{c}_{14}}{\tilde{c}_{44}} B_s~\sin3\theta \nonumber\\
\left[ \varepsilon_{\theta z} \right]_s&=& \frac{3}{2}\left[\frac{r_c^2}{r^2}-\frac{r_c^4}{r^4}+\eta^2 (1-\eta) \frac{r^2}{r_c^2} \right] \frac{\tilde{c}_{14}}{\tilde{c}_{44}} B_s\cos3\theta \nonumber\\
\end{eqnarray}
where
\begin{eqnarray}
  &&\frac{\tilde{c}_{14}}{\tilde{c}_{44}} B_s=-\frac{f}{\sqrt{2}}\frac{c_{11}-c_{12}-2c_{44}}{c_{11}-c_{12}+c_{44}}~~\times\nonumber\\&&\frac{3(c_{11}+2c_{12})}{(c_{11}-c_{12}+4c_{44})[\eta+(1-\eta
)\chi]+(2c_{11}+4c_{12}+2c_{44})}\nonumber
\end{eqnarray}
\begin{figure}
\resizebox{0.5\textwidth}{!}{\includegraphics {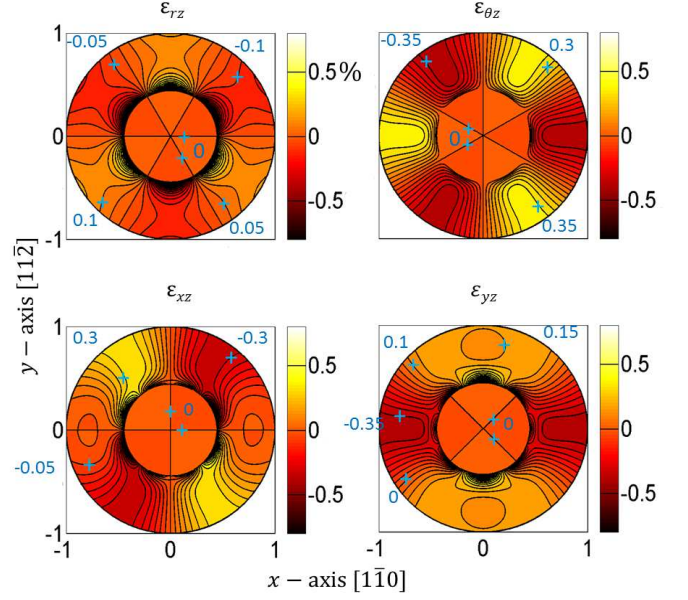}}
\caption{Strain map for an InAs core with an InP shell, with the
same area ratio $\eta=0.2$ as in ref.~\cite{Boxberg}. The lattice
mismatch is $f=-3.15\%$, stiffness ratio $\chi=1.2$, and
$B_s\tilde{c}_{14}/\tilde{c}_{44}=-0.82\%$. All contour line
spacings are $0.05\%$. The shear components $\varepsilon_{rz}$ and
$\varepsilon_{\theta z}$ reveal the trigonal symmetry. This symmetry
is masked if we plot $\varepsilon_{xz}$ or $\varepsilon_{xz}$ but
these plots can be compared qualitatively to ref.~\cite{Gronqvist},
and quantitatively (but taking into account the different axes used)
to ref.~\cite{Boxberg}. \label{shearStrainMap}}
\end{figure}
The results are shown in fig.~\ref{axialShear} for an InAs-InP NW
with the area ratio of ref.~\cite{Boxberg}, and for a thick shell.
Maps are shown in fig.~\ref{shearStrainMap}. Note the discontinuity
of $\varepsilon_{rz}$ at the interface, and its fast decay while
$\varepsilon_{\theta z}$ progressively increases from zero and stays
finite far into the shell. There is a complete agreement with the
results of numerical calculations in ref.~\cite{Boxberg}.

Apart from the presence of this additional shear strain, the other
strain components are modified by terms of the order of
$(\tilde{c}_{14}/\tilde{c}_{11})^2$. Taking again GaAs parameters,
we find that these second order terms are of the order of 1\%$\times
f$. As a result, the  change of the core strain induced by the
$\tilde{c}_{14}$ terms is negligible. Note also that the
contribution of the additional shear strain to $\sigma_{zz}$
vanishes due to the $\cos3\theta $ and $\sin3\theta $ factors.
\subsection{Summary and electronic properties}
To sum up, the strain configuration in a zinc-blende NW grown along
the $<111>$ axis is described by a cylindrical strain,
Eq.~\ref{strainhexa} and \ref{para111} (or
\ref{strainhexaSameStiffness} and \ref{strain111identical} if
$\chi=1$), complemented by an axial shear strain ("warping"),
Eq.~\ref{trigonalStrain}.

The Bir-Pikus Hamiltonian describing the coupling of holes to strain
has the same symmetry as the Luttinger Hamiltonian. When expressed
in the present trigonal basis (hole states $|\frac{3}{2}\rangle$,
$|\frac{1}{2}\rangle$, $|-\frac{1}{2}\rangle$ and $|-
\frac{3}{2}\rangle$ quantized along $[111]$, and strain tensor using
the axes $x=[1\bar{1}0]$, y=$[11\bar{2}]$, z=$[111]$), using the
symmetry arguments of ref.~\cite{Altarelli} as described in
ref.~\cite{Fishman}, the Hamiltonian writes
\begin{equation}\label{BirPikus111}
    \begin{pmatrix}
    P+Q & -S & R & 0 \\
    -S^* & P-Q & 0 & R \\
    R^* & 0 & P-Q & S \\
    0 & R^* & S^* & P+Q \\
    \end{pmatrix}
\end{equation}
with
\begin{eqnarray}
  P &=& -a(\varepsilon_{xx}+\varepsilon_{yy}+\varepsilon_{zz})\nonumber  \\
  Q &=& \frac{d}{2 \sqrt{3}}(\varepsilon_{xx}+\varepsilon_{yy}-2\varepsilon_{zz})\nonumber  \\
  R &=& -\frac{\sqrt{3}}{6}(b+\frac{2d}{\sqrt{3}}) (\varepsilon_{xx}-\varepsilon_{yy}-2i\varepsilon_{xy})\nonumber  \\&&+\frac{2}{\sqrt{6}}(b-\frac{d}{\sqrt{3}}) (\varepsilon_{xz}+i\varepsilon_{yz})\nonumber \\
  S &=& \frac{\sqrt{3}}{3}(2b+\frac{d}{\sqrt{3}} ) (\varepsilon_{xz}-i\varepsilon_{yz})\nonumber \\&&-\frac{1}{\sqrt{6}}(b-\frac{d}{\sqrt{3}}) (\varepsilon_{xx}-\varepsilon_{yy}+2i\varepsilon_{xy})\nonumber
\end{eqnarray}

In the core, apart from a small axial shear strain, which takes
non-vanishing values close to the interface but remains very small,
the strain comprises the hydrostatic strain
$(\varepsilon_{xx}+\varepsilon_{yy}+\varepsilon_{zz})$ and the
trigonal strain
$(\varepsilon_{xx}+\varepsilon_{yy}-2\varepsilon_{zz})$. The
Bir-Pikus Hamiltonian is diagonal in the trigonal basis, with a
splitting equal to $2Q$; the Luttinger Hamiltonian gives the
effective masses of the eigenstates: the mass along the NW axis
(determining the density of states and transport properties) is
$m^*=m_0/(\gamma_1 - 2\gamma_3)$ for the $|\pm \frac{3}{2}\rangle$
holes, and $m^*=m_0/(\gamma_1 + 2\gamma_3)$ for the $|\pm
\frac{1}{2}\rangle$ holes, the mass in the plane (governing
confinement) being $m^*=m_0/(\gamma_1\pm\gamma_3)$. This was used in
section \ref{exciton}.

In the shell, close to the interface, the dominant contribution is
the shear strain with cylindrical symmetry: it adds non-diagonal
matrix elements ($R$ and $S$) to the Bir-Pikus Hamiltonian, which
mixes the previous states. As in the previous case of a wurtzite NW,
half of the holes are confined to the interface. If we consider the
whole NW, the axial symmetry is preserved, so that the eigenstates
in the core retain their symmetry, with some mixing expected to take
place in narrow NWs. However, there is also a contribution from the
warping terms in the shell, which adds a modulation with a 3-fold
symmetry to the hole potential: this complex structure may
contribute to localization, particularly in NWs with a thick shell.

This deformation potential landscape is complemented by the
piezoelectric effect \cite{Boxberg}. Again, the polarization in the
core is along the axis, determined by $-e_{14}
(\varepsilon_{xx}+\varepsilon_{yy}-2\varepsilon_{zz})/\sqrt{3}$
where $e_{14}$ is the unique coefficient of the piezoelectric tensor
(the indices refer to the cubic axes). A complex lanscape however
emerges in the shell from the presence of in-plane and axial shear
strains, and of additional terms in the piezoelectric tensor written
in the trigonal axes \cite{BoxbergNL,Boxberg,Schulz}.
\section{Cubic semiconductors along $<001>$}
By contrast to the (111) plane of the zinc-blende structure, which
is quite isotropic, the (001) plane is known to be strongly
anisotropic. This is obvious on the stiffness tensor written in the
$\textbf{e}_r$, $\textbf{e}_{\theta}$, $\textbf{e}_z$ axes, obtained
by rotating the cubic axes by an angle $\theta)$ around $z$
(\emph{i.e.}, it is written in cylindrical coordinates):

\begin{equation}\label{stiffnessZB001Rotated}
\begin{pmatrix}
\hat{c}_{11}+\frac{c \cos 4\theta }{4}&\hat{c}_{12}-\frac{c \cos 4\theta }{4}&\hat{c}_{13}&0&0&0 \\
\hat{c}_{12}-\frac{c \cos 4\theta }{4}&\hat{c}_{11}+\frac{c \cos 4\theta }{4}&\hat{c}_{13}&0&0&0 \\
\hat{c}_{13}&\hat{c}_{13}&\hat{c}_{33}&0&0&0 \\
0&0&0&\hat{c}_{44}&0&0 \\
0&0&0&0&\hat{c}_{44}&0 \\
0&0&0&0&0&\hat{c}_{66}-\frac{c \cos 4\theta }{4} \\
\end{pmatrix}\nonumber \\
\end{equation}

with
\begin{eqnarray}\label{stiffnessZB001cylind}
&&\hat{c}_{11}=\frac{3c_{11}+c_{12}+2c_{44}}{4}\nonumber \\
&&\hat{c}_{12}=\frac{c_{11}+3c_{12}-2c_{44}}{4}\nonumber \\
&&\hat{c}_{66}=\frac{c_{11}-c_{12}+2c_{44}}{4}\nonumber\\
&&\hat{c}_{13}=c_{12},~ \hat{c}_{33}=c_{11},~ \hat{c}_{44}=c_{44}.
\end{eqnarray}
Ref.~\cite{Trammel} proposes a solution where one assumes the
cylindrical form of the displacement field, $u_r(r)=Ar+B/r$, and
writes boundary conditions at the interfaces/surfaces only in the
cubic directions. Actually, as we show now, the cylindrical
displacement field is not a solution of the Lam\'{e} - Clapeyron -
Navier equation, and the boundary conditions are not valid for other
directions of the basal plane. We thus propose a solution in two
steps, along the line we followed in the previous section for the NW
with trigonal axis. We thus identify the stiffness constants which
give the better approximation by a cylindrical solution, and we
calculate the additional strain with four-fold symmetry, which is
now a generalized in-plane shear strain.

\begin{figure}
\resizebox{0.4\textwidth}{!}{\includegraphics {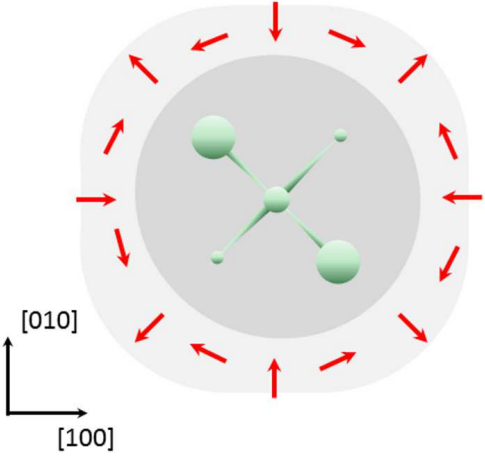}}
\caption{In-plane body forces in the shell of a $<001>$ NW, with
$f<0$, due to the four-fold symmetry of the zinc-blende structure.
The tetrahedron of the atomic structure is shown in
green.\label{excessStrain3}}
\end{figure}
The present form of the stiffness tensor,
Eq.~\ref{stiffnessZB001cylind}, identifies two contributions:
\begin{itemize}
  \item one with cylindrical symmetry (that with the $\hat{c}$, note that
$\hat{c}_{11}-\hat{c}_{12}=2 \hat{c}_{66}$); if we keep only this
contribution, the strain configuration is that of
Eq.~\ref{strainhexa} or Eq.~\ref{strainhexaSameStiffness}, where the
$c_{ij}$'s are replaced by the $\hat{c}_{ij}$'s;
  \item  one, proportional to $c$, with the
expected fourfold symmetry around $z$. As mentioned earlier, with
$c<0$, a zinc-blende crystal is softer against a pure tetragonal
stress (along a cubic axis, $\cos 4\theta =1$) than against any
other stress, in particular along a $<110>$ axis ($\cos 4\theta
=-1$).
\end{itemize}
Omitting the terms violating the translation invariance  and
identifying cylindrical contributions (in $\hat{c}$) and
contributions due to the cubic anisotropy (proportional to $c$), we
obtain for the Lam\'{e} - Clapeyron - Navier equation in cartesian
coordinates:
\begin{eqnarray} \label{LCN001cut}
    &&\frac{\hat{c}_{11}-\hat{c}_{12}}{2} (\frac{\partial ^2}{\partial x^2}+\frac{\partial ^2}{\partial y^2})u_x +\frac{\hat{c}_{11}+\hat{c}_{12}}{2}   \frac{\partial}{\partial x}(\frac{\partial u_x}{ \partial x}+\frac{\partial u_y}{ \partial y})\nonumber \\
    &&~~~~~~~~~~~~+\frac{c}{4}(\frac{\partial ^2u_x}{\partial x^2}-\frac{\partial ^2u_x}{\partial y^2}-2\frac{\partial ^2u_y}{\partial x \partial y})=0\nonumber\\
    &&\frac{\hat{c}_{11}-\hat{c}_{12}}{2} (\frac{\partial ^2}{\partial x^2}+\frac{\partial ^2}{\partial y^2})u_y +\frac{\hat{c}_{11}+\hat{c}_{12}}{2}   \frac{\partial}{\partial y}(\frac{\partial u_x}{ \partial x}+\frac{\partial u_y}{ \partial y})\nonumber \\
    &&~~~~~~~~~~~~+\frac{c}{4}(\frac{\partial ^2u_y}{\partial y^2}-\frac{\partial ^2u_y}{\partial x^2}-2\frac{\partial ^2u_x}{\partial x \partial y})=0\nonumber\\
    &&\frac{\partial ^2u_z}{\partial x^2}+\frac{\partial ^2u_z}{\partial y^2}=0
\end{eqnarray}
Inserting the cylindrical solution $u_r(r)=Ar+B/r$ reveals
non-vanishing contributions from the terms proportional to $c$. As
in the previous case, in a calculation to first order in $c$, these
terms act as body forces and generate an additional displacement
field $\delta \textbf{u}$, proportional to $c$. Even if these terms
look quite similar to those already encountered for the $<111>$ NW
(they amount to $8Br_c^2\sin 3\theta/r^3$ for the first equation of
Eq.~\ref{LCN001cut} and $-8Br_c^2 \cos 3\theta /r^3$ for the second
one), they appear as body forces in the basal plane, organized as a
\emph{transverse shear stress} \cite{Tsukrov} \emph{with four-fold
symmetry}:
\begin{eqnarray}\label{LCN2D001}
    F_r&=&c\frac{2B_sr_c^2}{r^3}\cos 4\theta  \nonumber\\
    F_\theta&=&-c\frac{2B_sr_c^2}{r^3}\sin 4\theta
\end{eqnarray}
in the shell, and zero in the core (fig.~\ref{excessStrain3}). We
thus have to find an additional in-plane displacement $\delta
\textbf{u} (r,\theta)$ which is the response of the transversely
isotropic system to these forces. The relevant part of the
Lam\'{e}-Clapeyron-Navier equation is a two-dimensional equation:
\begin{equation}\label{}
   \frac{\hat{c}_{11}-\hat{c}_{12}}{2}\triangle~\delta
    \textbf{u}+\frac{\hat{c}_{11}+\hat{c}_{12}}{2}\nabla(\nabla.\delta
    \textbf{u})=\textbf{F} \nonumber
\end{equation}
or, defining a Poisson ratio
$\nu=\hat{c}_{12}/(\hat{c}_{11}+\hat{c}_{12})=(c_{11}+3c_{12}-2c_{44})/4(c_{11}+c_{12})$,
\begin{equation}\label{LCN001Poisson}
   (1-2\nu)\triangle~\delta
    \textbf{u}+\nabla(\nabla.\delta
    \textbf{u})=\frac{2\textbf{F}}{\hat{c}_{11}+\hat{c}_{12}}
\end{equation}

The solution is
\begin{eqnarray}\label{}
\frac{\delta u_r}{r_c}=\frac{c}{\hat{c}_{11}+\hat{c}_{12}}B_s g_r(r)
\cos 4\theta \nonumber \\
\frac{\delta u_\theta}{r_c}=\frac{c}{\hat{c}_{11}+\hat{c}_{12}}B_s
g_\theta(r) \sin 4\theta
\end{eqnarray}
where $g_r$ and $g_\theta$ are two dimensionless functions of
$r/r_c$ - more precisely they are sums of five terms in $(r/r_c)^n$
with $n=-1, \pm 3, \pm5$, which are given in appendix
\ref{generalShear}.

\textbf{To sum up,} the strain configuration in a core-shell NW
grown along $<001>$ is given by Eq.~\ref{strainhexa}, with
\begin{eqnarray} \label{para001}
B_s&=&-\frac{c_{11}+2c_{12}}{(\frac{c_{11}-c_{12}}{2}+c_{44})[\eta+(1-\eta
)\chi]+(c_{11}+c_{12})}f \nonumber \\
f_{\bot}+B_s&=&\frac{(\frac{c_{11}-c_{12}}{2}+c_{44})[\eta+(1-\eta
)\chi]-c_{12}}{(\frac{c_{11}-c_{12}}{2}+c_{44})[\eta+(1-\eta
)\chi]+(c_{11}+c_{12})}f\nonumber \\
\end{eqnarray}
If the materials have the same hardness ($\chi=1$), this reduces to
Eq.~\ref{strainhexaSameStiffness} and
\begin{eqnarray} \label{paraIso001}
B_s&=&-\frac{2(c_{11}+2c_{12})}{3c_{11}+c_{12}+2c_{44}}f \nonumber \\
f_{\bot}+B_s&=&\frac{c_{11}-3c_{12}+2c_{44}}{3c_{11}+c_{12}+2c_{44}}f\nonumber \\
\end{eqnarray}
This is complemented by an in-plane shear strain which writes (for
$\chi=1$):
\begin{eqnarray} \label{para001}
\varepsilon_{rr}&=&\frac{c}{c_{11}+c_{12}} B_s \cos 4\theta ~g_{rr}(\frac{r}{r_c})   \nonumber \\
\varepsilon_{r\theta}&=&\frac{c}{c_{11}+c_{12}} B_s \cos 4\theta ~g_{r\theta}(\frac{r}{r_c}) \nonumber \\
\varepsilon_{\theta \theta}&=&\frac{c}{c_{11}+c_{12}} B_s \sin
4\theta ~g_{\theta \theta}(\frac{r}{r_c}) \nonumber
\end{eqnarray}
where $g_{\theta \theta}=4 g_{\theta}+g_r$, and $g_r$, $g_\theta$,
$g_{r\theta}$ and $g_{rr}$ are given in Appendix \ref{generalShear},
Eq.~\ref{transfer001} with the coefficients given in \ref{001alphas}
for the shell and the core.

\begin{figure}
\resizebox{0.5\textwidth}{!}{\includegraphics {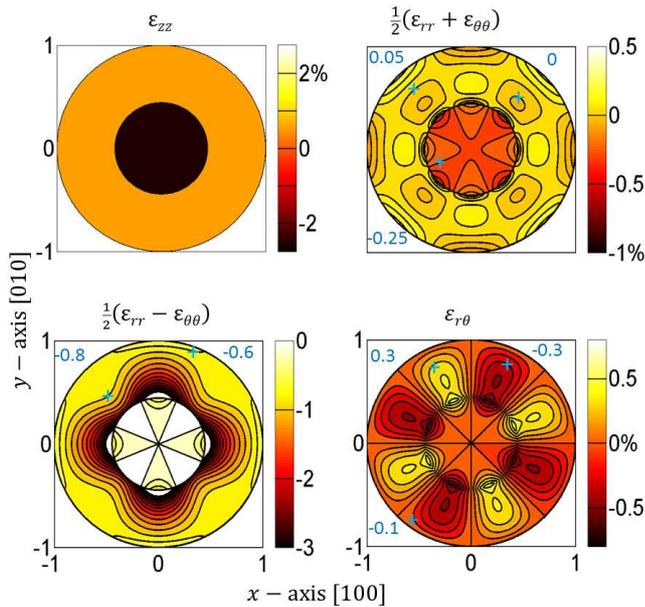}}
\caption{Strain maps for an InAs core with an InP shell, with the
same area ratio $\eta=0.2$ as in ref.~\cite{Boxberg}. The NW axis is
$<001>$ and we used $\chi=1$. The mismatch is $f=-3.15 \%$ and $B_sc
/(c_{11}+c_{12})=-0.87\%$. The axial strain component
$\varepsilon_{zz}$ is uniform. The contour line spacing is $0.05\%$
for $\frac{1}{2}(\varepsilon_{rr}+\varepsilon_{\theta \theta})$,
$0.2\%$ for $\frac{1}{2}(\varepsilon_{rr}-\varepsilon_{\theta
\theta})$, and $0.1\%$ for $\varepsilon_{r\theta}$ All in-plane
strain components exhibit a four-fold contribution due to the
crystal anisotropy. \label{StrainMap001}}
\end{figure}

\begin{figure}
\resizebox{0.5\textwidth}{!}{\includegraphics {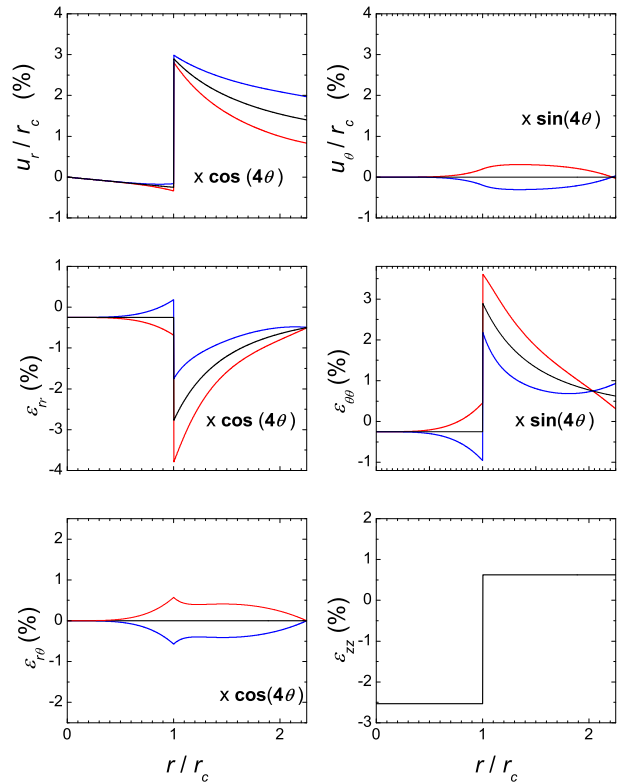}}
\caption{Radial profile of the displacement field and the strain
components for the same NW as in fig.~\ref{StrainMap001}. The NW
axis is $<001>$. The (black) central lines are the cylindrical
contribution, the two other lines show the extreme values due to the
cubic anisotropy, with the dependence on polar angle as indicated
(blue: $\sin 4\theta $ or $\cos 4\theta =1$; red: $\sin 4\theta $ or
$\cos 4\theta =-1$). \label{001profiles}}
\end{figure}

 Fig.~\ref{StrainMap001} shows the strain map for an
InAs-InP NW with the same area ratio $\eta$ as in
fig.~\ref{isoStrainMap} and \ref{shearStrainMap}, and
ref.~\cite{Boxberg}, but with the NW axis along $<001>$.
Fig.~\ref{001profiles} displays the radial profiles of the in-plane
displacement field, the in-plane strain components (the cylindrical
contribution and the modulation in $\sin 4\theta $ or $\cos 4\theta
$ due to cubic anisotropy), and the axial strain. The cubic
contribution is negligible in the central part of the core, and
remains small close to the interface; in the shell, it takes
significant values, yet smaller than the cylindrical contribution.
Further contributions should bring terms of higher order in $4
\theta$, with the order of magnitude of the second order terms
around $c/4c_{11}$, \emph{i.e.}, again, a few \% in GaAs.
\section{Discussion and conclusion}
The present study proposes an analytical treatment of the strain
distribution in core-shell and multishell NWs with circular section.
Several comparisons have been given with numerical treatments using
either a valence force field model
\cite{Niquet,BoxbergNL,Gronqvist,Herstroffer,Boxberg,Hocevar}, or a
finite element implementation of continuum elasticity
\cite{Boxberg}. Even if commercial packages now exist which will
give the same results as these numerical treatments, the analytical
treatment remains faster, and it favors a more comprehensive
understanding.

It has been recognized for a long time \cite{Wortman} that the (111)
plane of a cubic crystal (in the present case, zinc-blende or
diamond semiconductor) is isotropic - and the same property also
holds for the ($a,b$) plane of the wurtzite structure.

In a core-shell NW, this remains valid for a core-shell NW oriented
along the $c$-axis of the wurtzite structure. This transverse
isotropy has several consequences which are reminiscent from the
case of a fully isotropic material.
\begin{itemize}
  \item The longitudinal strain is decoupled from the in-plane
  strain. It is uniform in the core and in the shell, and results from
  a sharing of the lattice mismatch along the $c$ axis,
  inversely proportional to the cross section areas.
  \item The in-plane strain in the core is isotropic and uniform. It
  is the result of a partial compensation between the direct effect
  of the in-plane lattice mismatch (the $c_{13}f_{\bot}$ contribution in eq.
  \ref{strainhexa}) and the Poisson effect from the longitudinal mismatch (the $(c_{11}-c_{12})f_{\|}$
  contribution). The simple result obtained for a fully
  isotropic material (a factor of $-\frac{1}{2}$ in the (shear strain)
  / (isotropic strain) ratio when comparing the
  NW to the thin layer) must be adapted to the relevant stiffness
  constants. In the case of a GaN-AlN NW, the compensation is
  reinforced by the different values of the lattice mismatch in the
  two directions, so that the in-plane strain in the core is reduced
  by one order of magnitude.
  \item Actually the main part of the in-plane lattice mismatch is
  accommodated by the in-plane shear strain, which rotates around
  the  interface so that the circular symmetry is maintained. The
  fact that this strain is restricted to the vicinity of the
  interface is a consequence of the Saint-Venant principle.
\end{itemize}

It is interesting to note that this shear strain induces a potential
which can be used to confine holes in the shell in the vicinity of
the interface, far from the sidewall. It thus allows the design of
type-II core-shell NWs where both the electrons (in the core) and
the holes (in the shell) are kept away from surface defects.

The strain distribution in a NW oriented along the $<111>$ axis of a
semiconductor with the zinc-blende (or diamond) structure is more
complex. Shear strains and shear stresses are expected, and they
appear in the numerical studies. They are due to the trigonal
symmetry around the $<111>$ axis, and more precisely to the presence
of tetrahedral building blocks with a single orientation - while two
orientations co-exist in the wurtzite structure \cite{Martin}. The
present analysis shows that these shear strain indeed exist in the
shell, and that their influence on the strain in the core is small.
The uniform  strain, isotropic in the plane, which exists in the
core can be calculated analytically using the stiffness tensor
appropriate for the $<111>$ orientation.

The same method gives analytical results also in the case of a NW
with the zinc-blende (or diamond) structure grown along a cubic
axis: then in-plane strain with four-fold symmetry develops in
addition to the cylindrical configuration.

Note that with these two examples (NWs grown along the trigonal or
along the cubic axis of the zinc-blende / diamond structure), we
obtain the two possible types of additional generalized shear strain
(axial or in-plane). NWs with other types of symmetry are expected
to involve combinations of these two types of generalized
shear-strain.

While the present study assumes a circular basis of the NWs,
numerical studies also reveals the role of facets: for a hexagonal
basis, the strain in the core is not uniform in the corners of the
hexagons. An analytical method has been proposed for isotropic
materials in ref.~\cite{Faux}. Alternately, a possible extension of
the present method would be to express the difference between the NW
with a polygonal section and that with a circular one, as a field of
body forces, which would be localized at the corners of the polygon;
then, as we did for the crystalline anisotropy, we could calculate
the response of the system to that field. Nevertheless, the
comparison between the present calculation and the plateaus values
from numerical studies suggests again a quantitative agreement,
which can be seen as another consequence of the Saint-Venant's
principle.

Finally, multishell NWs are currently proposed for applications such
as the direct-bandgap emission from $<001>$ Si-Ge NWs \cite{Zhang},
or a reduction of piezoelectric effects in wurtzite or $<111>$
zinc-blende NWs \cite{Eymery}. The present study shows that a shear
strain exists in such lateral QWs, different from well to well. The
transfer matrix method can also be used to incorporate the effects
of surface stress, which may become significant in narrow NWs
\cite{Schmidt}, or of surface layers (oxide for instance), two
effects which will be difficult to disentangle.

This work was done in the joint CNRS-CEA group "Nanophysique \&
semiconducteurs", and in the frame of the ANR project "Magwires"
(ANR-11-BS10-013). We thank Yann-Michel Niquet, Mo\"{\i}ra Hocevar
and all the members of the Magwires project for many discussions and
for communicating their results.

\appendix
\numberwithin{equation}{section}
\section {Lam\'{e} - Clapeyron -
Navier equations} \label{completeLCN}

The full Lam\'{e} - Clapeyron - Navier equations are written for the
three crystal structures and orientations.

\subsection{Wurzite, $c$-axis}

\begin{eqnarray} \label{LCNhexafull}
    &&c_{11}\frac{\partial ^2u_x}{\partial x^2}+\frac{c_{11}-c_{12}}{2}\frac{\partial ^2u_x}{\partial y^2}+c_{44}\frac{\partial ^2u_x}{\partial z^2}\nonumber \\&&+\frac{c_{11}+c_{12}}{2}\frac{\partial ^2u_y}{\partial x \partial y}+(c_{13}+c_{44})\frac{\partial ^2u_z}{\partial x \partial z}=0\nonumber \\
    &&\frac{c_{11}-c_{12}}{2}\frac{\partial ^2u_y}{\partial x^2}+c_{11}\frac{\partial ^2u_y}{\partial y^2}+c_{44}\frac{\partial ^2u_y}{\partial z^2}\nonumber \\&&+(c_{13}+c_{44})\frac{\partial ^2u_z}{\partial y \partial z}+\frac{c_{11}+c_{12}}{2}\frac{\partial ^2u_x}{\partial x \partial y}=0\nonumber\\
    &&c_{44}\frac{\partial ^2u_z}{\partial x^2}+c_{44}\frac{\partial ^2u_z}{\partial y^2}+c_{33}\frac{\partial ^2u_z}{\partial z^2}\nonumber \\&&+(c_{13}+c_{44})\frac{\partial ^2u_x}{\partial x \partial z}+(c_{13}+c_{44})\frac{\partial ^2u_y}{\partial y \partial z}=0\nonumber\\
\end{eqnarray}

\subsection{Zinc-blende, $<111>$ axis}

\begin{eqnarray} \label{LCN111full}
    &&\tilde{c}_{11}\frac{\partial ^2u_x}{\partial x^2}+\frac{\tilde{c}_{11}-\tilde{c}_{12}}{2}\frac{\partial ^2u_x}{\partial y^2}+\tilde{c}_{44}\frac{\partial ^2u_x}{\partial z^2}\nonumber \\&&+\frac{\tilde{c}_{11}+\tilde{c}_{12}}{2}\frac{\partial ^2u_y}{\partial x \partial y}+(\tilde{c}_{13}+\tilde{c}_{44})\frac{\partial ^2u_z}{\partial x \partial z}\nonumber \\
    &&+2\tilde{c}_{14}\frac{\partial ^2u_x}{\partial y \partial z}+2\tilde{c}_{14}\frac{\partial ^2u_y}{\partial z \partial x}+2\tilde{c}_{14}\frac{\partial ^2u_z}{\partial x \partial y}=0\nonumber \\
    &&\frac{\tilde{c}_{11}-\tilde{c}_{12}}{2}\frac{\partial ^2u_y}{\partial x^2}+\tilde{c}_{11}\frac{\partial ^2u_y}{\partial y^2}+\tilde{c}_{44}\frac{\partial ^2u_y}{\partial z^2}\nonumber \\&&+(\tilde{c}_{13}+\tilde{c}_{44})\frac{\partial ^2u_z}{\partial y \partial z}+\frac{\tilde{c}_{11}+\tilde{c}_{12}}{2}\frac{\partial ^2u_x}{\partial x \partial y}\nonumber\\
    &&+2\tilde{c}_{14}\frac{\partial ^2u_x}{\partial z \partial x}+\tilde{c}_{14}\frac{\partial ^2u_z}{\partial x^2}-2\tilde{c}_{14}\frac{\partial ^2u_y}{\partial y \partial z}-\tilde{c}_{14}\frac{\partial ^2u_z}{\partial y^2}=0\nonumber\\
    &&\tilde{c}_{44}\frac{\partial ^2u_z}{\partial x^2}+\tilde{c}_{44}\frac{\partial ^2u_z}{\partial y^2}+\tilde{c}_{33}\frac{\partial ^2u_z}{\partial z^2}\nonumber \\&&+(\tilde{c}_{13}+\tilde{c}_{44})\frac{\partial ^2u_x}{\partial x \partial z}+(\tilde{c}_{13}+\tilde{c}_{44})\frac{\partial ^2u_y}{\partial y \partial z}\nonumber \\&&+2\tilde{c}_{14}\frac{\partial ^2u_x}{\partial x \partial y}+\tilde{c}_{14}\frac{\partial ^2u_y}{\partial x^2}-\tilde{c}_{14}\frac{\partial ^2u_y}{\partial y^2}=0\nonumber\\
\end{eqnarray}

\subsection{Zinc-blende, $<001>$ axis}

\begin{eqnarray} \label{LCN001}
    &&c_{11}\frac{\partial ^2u_x}{\partial x^2}+c_{44}\frac{\partial ^2u_x}{\partial y^2}+c_{44}\frac{\partial ^2u_x}{\partial z^2}\nonumber \\&&+(c_{12}+c_{44})\frac{\partial ^2u_y}{\partial x \partial y}+(c_{12}+c_{44})\frac{\partial ^2u_z}{\partial x \partial z}=0\nonumber \\
    &&c_{44}\frac{\partial ^2u_y}{\partial x^2}+c_{11}\frac{\partial ^2u_y}{\partial y^2}+c_{44}\frac{\partial ^2u_y}{\partial z^2}\nonumber \\&&+(c_{12}+c_{44})\frac{\partial ^2u_z}{\partial y \partial z}+(c_{12}+c_{44})\frac{\partial ^2u_x}{\partial x \partial y}=0\nonumber\\
    &&c_{44}\frac{\partial ^2u_z}{\partial x^2}+c_{44}\frac{\partial ^2u_z}{\partial y^2}+c_{11}\frac{\partial ^2u_z}{\partial z^2}\nonumber \\&&+(c_{12}+c_{44})\frac{\partial ^2u_x}{\partial x \partial z}+(c_{12}+c_{44})\frac{\partial ^2u_y}{\partial y \partial z}=0\nonumber\\
\end{eqnarray}

\section {Generalized shear
strains} \label{generalShear}

The present study involves Lam\'{e} - Clapeyron - Navier equations
describing the response of a system which is invariant under a
translation along the $z$-axis and isotropic in the basal $xy$
plane, to body forces which are periodic in a rotation around the
$z$-axis: $F_z=F \sin 3\theta $ for a zinc-blende NW along a $<111>$
axis, and $F_r=F \cos 4\theta $, $F_\theta=-F \sin 4\theta $ for a
zinc-blende NW along a $<001>$ axis. Note that $F_z=F \cos (
\theta)$ over the whole structure describes a uniform axial shear
strain applied to the system, and $F_r=F \sin 2\theta$, $F_\theta=F
\cos 2\theta$ a uniform transverse shear strain: a transfer matrix
method was proposed in ref.~\cite{Tsukrov} for multishell NWs
submitted to these two types of shear strain. The present study
involves similar body forces distributions with a faster dependence
on $\theta$, localized in the shell: $F_z=F \sin p\theta$ with
$p=3$, and $F_r=F \cos p\theta$, $F_\theta=-F \sin p\theta$ with
$p=4$. Other orientations of the NWs will involve combinations of
such body forces distributions.

We thus have to calculate a displacement field $\delta \textbf{u}$,
solution of the Lam\'{e} - Clapeyron - Navier equation
\begin{eqnarray} \label{LCNforces}
    &&\frac{\hat{c}_{11}-\hat{c}_{12}}{2} \left(\frac{\partial ^2}{\partial x^2}+\frac{\partial ^2}{\partial y^2}\right)\delta u_x +\frac{\hat{c}_{11}+\hat{c}_{12}}{2}   \frac{\partial}{\partial x}\left(\frac{\partial \delta u_x}{ \partial x}+\frac{\partial \delta u_y}{ \partial y}\right)\nonumber \\
    &&~~~~~~~~~~~~+F_x=0\nonumber\\
    &&\frac{\hat{c}_{11}-\hat{c}_{12}}{2} \left(\frac{\partial ^2}{\partial x^2}+\frac{\partial ^2}{\partial y^2}\right)\delta u_y +\frac{\hat{c}_{11}+\hat{c}_{12}}{2}   \frac{\partial}{\partial y}\left(\frac{\partial \delta u_x}{ \partial x}+\frac{\partial \delta u_y}{ \partial y}\right)\nonumber \\
    &&~~~~~~~~~~~~+F_y=0\nonumber\\
    &&\hat{c}_{44} \left(\frac{\partial ^2}{\partial x^2}+\frac{\partial ^2}{\partial y^2}\right)\delta u_z+F_z=0
\end{eqnarray}

As the response of a linear, transversely isotropic system to an
oscillating perturbation, the general solution is expected to show
the same oscillatory behavior, in $\cos (p\theta)$ or $\sin
p\theta$.

The boundary conditions are the continuity of the total displacement
field, $\textbf{u}+\delta \textbf{u}$, at the interface, and that
the stress components acting on the interface and on the sidewall
surface ($\sigma_{rr}$, $\sigma_{r\theta}$, $\sigma_{rz}$) all
vanish. The last condition must be achieved for the total stress,
corresponding to $\textbf{u}+\delta \textbf{u}$. For the
displacement field, it is sufficient to write that the additional
displacement field does not break the contact which has been
established by the cylindrical displacement field, hence $\delta
\textbf{u}=0$. Note that the symmetry of the system and that of the
shear strain strongly reduce the number of parameters to be
determined from boundary conditions. For instance, the condition
that the integral of $\sigma_{zz}$ vanishes is automatically
preserved by the oscillating character of $\delta \textbf{u}$.

\subsection{Axial shear strain and $<111>$ NWs}
In the absence of driving force in the basal plane, we keep $\delta
u_r=0$ and $\delta u_\theta=0$, and look for $\delta
u_z=\varphi(r)\sin p\theta$, with $\varphi(r)$ obeying
eq.~\ref{trigoStrain}. In cylindrical coordinates, that reads
\begin{equation}\label{}
    \tilde{c}_{44} \left(\frac{\partial ^2}{\partial r^2}+\frac{1}{r} \frac{\partial}{\partial r}+\frac{1}{r^2}\frac{\partial ^2}{\partial \theta^2}\right)\left [\varphi (r) \sin p \theta\right
    ]+F_z=0 \nonumber
\end{equation}
or
\begin{equation}\label{}
    \tilde{c}_{44} \left(\frac{d ^2}{d r^2}+\frac{1}{r} \frac{d}{d r}-\frac{p^2}{r^2}\right)\varphi(r)~~\sin p \theta+F_z=0 \nonumber
\end{equation}
The general solution is the sum of functions $\sim r^n$: $n=-1$
provides a particular solution which compensates for $F_z$, and for
$n=\pm p$, the sum of derivatives vanishes.

With $F_z=\tilde{c}_{14} 8 B_s r_c^2 \sin 3\theta  /r^3$ in the
shell, we obtain
\begin{equation}\label{}
    \frac{\delta u_z}{r_c}= \left[ \alpha_3
    \rho^3+\alpha_{-3}
\rho^{-3}+\alpha_{-1} \rho^{-1} \right]
\frac{\tilde{c}_{14}}{\tilde{c}_{44}}  B_s \sin 3\theta  \nonumber
\end{equation}
where $\rho=r/r_c$, with $\alpha_{-1}^s=-1$ in the shell and
$\alpha_{-1}^c=0$ in the core. Also, $\alpha_{-3}^c=0$ in the core
to avoid a singularity at $r=0$. The additional strain is thus
\begin{eqnarray}\label{}
    &&\delta\varepsilon_{rz}=\frac{1}{2}\frac{ \partial}{\partial r}\delta
    u_z\nonumber \\
   && = \left[ 3\alpha_3 \rho^2-3\alpha_{-3}
\rho^{-4}-\alpha_{-1} \rho^{-2} \right]
\frac{\tilde{c}_{14}}{\tilde{c}_{44}}  B_s \sin 3\theta \nonumber
\end{eqnarray}
and
\begin{eqnarray}\label{}
    &&\sigma_{rz}=2\tilde{c}_{44} \delta\varepsilon_{rz}+ \tilde{c}_{14} \sin 3\theta (\epsilon_{rr}-\varepsilon_{\theta \theta })\nonumber \\
   && = \left[ 3 \alpha_3 \rho^2-3\alpha_{-3}
\rho^{-4}+(2-\alpha_{-1}) \rho^{-2} \right] \tilde{c}_{14} B_s \sin
3\theta \nonumber
\end{eqnarray}
The three remaining parameters $\alpha_3^c$, $\alpha_3^s$ and
$\alpha_{-3}^s$ are determined by the non-trivial boundary
conditions, on $u_z$ (at the interface) and $\sigma_{rz}$ (at the
interface and surface). It is quite convenient to write these
conditions using a transfer matrix:
\begin{eqnarray}\label{matrixdef}
\begin{pmatrix}{}
         (\frac{\delta u_z}{r_c})&\\
          (\frac{\sigma_{rz}}{\tilde{c}_{44}})& \\
       \end{pmatrix}
       &&= \frac{\tilde{c}_{14}}{\tilde{c}_{44}} B_s  \sin 3\theta
           \begin{pmatrix}{}
         \rho^3 & \rho^{-3} \\
         3\rho^2 & -3\rho^{-4} \\
       \end{pmatrix}
       \begin{pmatrix}{}
         \alpha_3 &\\
          \alpha_{-3} & \\
       \end{pmatrix}\nonumber \\
       &&+ \alpha_{-1} \frac{\tilde{c}_{14}}{\tilde{c}_{44}} B_s  \sin 3\theta
 \begin{pmatrix}{}
        \rho^{-1}&\\
        -3\rho^{-2}& \\
       \end{pmatrix}
\end{eqnarray}
At the interface ($\rho=1$), if we omit the difference in stiffness
coefficients between the two materials:
\begin{equation}\label{axialshearinterface}
\begin{pmatrix}
         1 & 1 \\
         3 & -3 \\
\end{pmatrix}
\begin{pmatrix}
         \alpha_3^c &\\
         0 & \\
\end{pmatrix}
=
\begin{pmatrix}
         1 & 1 \\
         3 & -3 \\
\end{pmatrix}
\begin{pmatrix}
         \alpha_3^s \\
         \alpha_{-3}^s\\
\end{pmatrix}
+
\begin{pmatrix}
         1& \\
         -3 &  \\
\end{pmatrix} \nonumber
\end{equation}
or
\begin{equation}\label{}
    \begin{pmatrix}
         \alpha_3^c &\\
         0 & \\
\end{pmatrix}
=
\begin{pmatrix}
         \alpha_3^s \\
         \alpha_{-3}^s\\
\end{pmatrix}
+
\begin{pmatrix}
         0 &\\
         1 & \\
\end{pmatrix} \nonumber
\end{equation}
At the surface, using eq.~\ref{matrixdef} at $r=r_s$
($\rho=1/\sqrt{\eta}$), and keeping only the second component of the
vectors (the stress which must be zero), we obtain
\begin{equation}\label{}
    0=(\alpha_3^s-\eta^3 \alpha_{-3}^s)-\eta^2 \nonumber
\end{equation}
Hence $\alpha_3^c=\alpha_3^s=\eta^2(1-\eta)$ and $\alpha_{-3}^s=-1$.

If we assume a different hardness with a single factor $\chi$
between the stiffness coefficients of the shell with respect to
those of the core material, eq.~\ref{axialshearinterface} becomes
\begin{equation}\label{axialshearinterface2}
\begin{pmatrix}
         1 & 1 \\
         3 & -3 \\
\end{pmatrix}
\begin{pmatrix}
         \alpha_3^c &\\
         0 & \\
\end{pmatrix}
=
\begin{pmatrix}
         1 & 1 \\
         3\chi & -3\chi \\
\end{pmatrix}
\begin{pmatrix}
         \alpha_3^s \\
         \alpha_{-3}^s\\
\end{pmatrix}
+
\begin{pmatrix}
         1& \\
         -3\chi &  \\
\end{pmatrix} \nonumber
\end{equation}
and the result is
\begin{eqnarray*}
  \alpha_3^c &=& \eta^2(1-\eta) \frac{2\chi}{1+\chi+\eta^3(1-\chi)}\nonumber\\
  \alpha_3^s &=& \eta^2(1-\eta) \frac{1+\chi}{1+\chi+\eta^3(1-\chi)} \nonumber\\
  \alpha_{-3}^s &=& -1-\eta^2(1-\eta)\frac{1-\chi}{1+\chi+\eta^3(1-\chi)}\nonumber
\end{eqnarray*}
The correction for $\chi$ non unity is small for the actual NW
configurations considered here: with $\chi=1.2$ and $\eta=0.2$, the
corrective factor is $10\%$ for $\alpha_3^c$ and negligible for the
shell.

This result was used in the case of the $<111>$ core-shell NWs and
it can be extended to multishell NWs.

\subsection{Transverse shear strain and $<001>$ NWs}
The problem is similar to the previous one: we have to find the
response of a system with transverse isotropic character, to a body
force distribution \textbf{F}. The body forces $F$ represent an
in-plane shear strain, with a four-fold symmetry due to the $\cos
4\theta$ factor. A usual shear strain would have a $\cos 2\theta$
and $\sin 2\theta$ factors, as described in ref.~\cite{Tsukrov}. The
solution is a bit more complex than the response to axial shear
because we are dealing with a 2D, not 1D, problem.

The equation to be solved, eq.~\ref{LCN001Poisson}, is, in polar
coordinates:
\begin{eqnarray}\label{LCN001polar}
   &&\left[ 2(1-\nu)\left(\frac{\partial^2}{\partial r^2}+\frac{1}{r}\frac{\partial}{\partial r}-\frac{1}{r^2}\right)+(1-2\nu)\frac{1}{r^2}\frac{\partial^2}{\partial \theta^2}\right] \delta
    u_r\nonumber\\
    &&+\left[\frac{1}{r}\frac{\partial^2}{\partial r \partial\theta}-(3-4\nu)\frac{1}{r^2}\frac{\partial}{\partial
    \theta}\right] \delta u_\theta
    +\frac{2F_r}{\hat{c}_{11}+\hat{c}_{12}}=0\nonumber\\
&&\left[\frac{1}{r}\frac{\partial^2}{\partial r
\partial\theta}+(3-4\nu)\frac{1}{r^2}\frac{\partial}{\partial
    \theta}\right] \delta u_r\nonumber\\
    &&+\left[(1-2\nu)\left(\frac{\partial^2}{\partial r^2}+\frac{1}{r}\frac{\partial}{\partial
    r}-\frac{1}{r^2}\right)+2(1-\nu)\frac{1}{r^2}\frac{\partial ^2}{\partial
    \theta^2}\right] \delta u_\theta\nonumber\\
    &&+\frac{2F_\theta}{\hat{c}_{11}+\hat{c}_{12}}=0
\end{eqnarray}

The general solution can be written
\begin{eqnarray}\label{}
\frac{\delta u_r}{r_c}=\frac{c}{\hat{c}_{11}+\hat{c}_{12}}B_s g_r(r)
\cos 4\theta \nonumber \\
\frac{\delta u_\theta}{r_c}=\frac{c}{\hat{c}_{11}+\hat{c}_{12}}B_s
g_\theta(r) \sin 4\theta
\end{eqnarray}
where $g_r$ and $g_\theta$ are two dimensionless functions which are
sums of terms in $\rho^n$ with $\rho=r/r_c$ and the $n$ are integer
(positive or negative).

For functions $\sim r^n \cos p\theta$ or $r^n \sin p\theta$, the
derivative contributions in eq.~\ref{LCN001polar} vanish if
$n^2=(p\pm 1)^2$. In the present case, $p=4$ hence $n=\pm3, \pm5$.
In addition, the prefactors $\alpha_n$ and $\alpha_n$ of $r^n$ for a
given value of $n$ are linked since the two equations of eq.
~\ref{LCN001polar} must be satisfied. Finally, the prefactors
$\alpha_{-1}$ and $\alpha'_{-1}$ are fully determined by the fact
that it is the $r^{-1}$ contribution in $g_r$ and $g_\theta$ which
makes eq. ~\ref{LCN001polar} to be satisfied. As the $\alpha_{n}$
with negative indices all vanish in the core (to avoid a singularity
at $r=0$), we have to determine six parameters, $\alpha^c_{3}$ and
$\alpha^c_{5}$ in the core, $\alpha^s_{3}$, $\alpha^s_{-3}$,
$\alpha^s_{5}$, and $\alpha^s_{-5}$ in the shell.

Boundary conditions are the continuity of $\delta\textbf{u}$ at the
interface, and the compensation of stress components acting on the
interface and sidewall surface. The relevant stress components are,
in the shell
\begin{eqnarray}\label{}
\sigma_{rr}&&=\hat{c}_{11} \delta \varepsilon_{rr}+ \hat{c}_{12} \delta \varepsilon_{\theta \theta}+\frac{c}{2} \cos 4\theta \frac{\varepsilon_{rr}-\varepsilon_{\theta \theta }}{2}\nonumber \\
&&=\hat{c}_{11} \frac{\partial }{\partial r} \delta u_r+
\hat{c}_{12} \left(\frac{\partial}{r\partial \theta} \delta u_\theta
+\frac{\delta u_r}{r}\right)-\frac{c}{2} B_s \rho^{-2} \cos
4\theta\nonumber
\\
\sigma_{r\theta}&&=\hat{c}_{66} \varepsilon_{r \theta}=\hat{c}_{66}
\frac{1}{2}\left( \frac{\partial}{r\partial \theta}  \delta
u_r+\frac{\partial}{\partial r} \delta u_\theta-\frac{ \delta
u_\theta}{r}\right)\nonumber
\end{eqnarray}
They are similar in the core, but for $B_c=0$.

That makes six boundary conditions.

Writing the two stress components
\begin{eqnarray}\label{}
 \frac{\sigma_{r \theta}}{\hat{c}_{66}}=\frac{c}{\hat{c}_{11}+\hat{c}_{12}}B_s
g_{r\theta}(r)
\cos 4\theta \nonumber \\
 \frac{\sigma_{r r}}{\hat{c}_{11}+\hat{c}_{12}}=\frac{c}{\hat{c}_{11}+\hat{c}_{12}}B_s
g_{rr}(r) \sin 4\theta
\end{eqnarray}

the four functions $g_r$, $g_\theta$, $g_{r\theta}$ and $g_{rr}$
which are submitted to boundary conditions at the interface can be
once again expressed in the frame of a transfer matrix treatment.
\begin{equation}\label{transfer001}
\begin{pmatrix}
        g_r \\
        g_\theta\\
        g_{r\theta} \\
        g_{rr}\\
\end{pmatrix}
=\textsf{\textbf{M}}\left(\rho\right)
\begin{pmatrix}
         \alpha_3 \\
         \alpha_{-3} \\
         \alpha_5 \\
         \alpha_{-5} \\
\end{pmatrix}
+ \mathbf{V}\left(\rho\right)
\end{equation}
where
\begin{equation}\label{}
\textsf{\textbf{M}}(1)=
\begin{pmatrix}
         1&(3-2\nu)&(1+2\nu)&1& \\
         -1&2\nu&-2(2-\nu)&1& \\
         -3&-6&-10&-5& \\
         3(1-2\nu)&-9(1-2\nu)&5(1-2\nu)&-5(1-2\nu)& \\
\end{pmatrix}\nonumber
\end{equation}
\begin{equation}\label{}
\textsf{\textbf{M}}(\rho)=
\begin{pmatrix}
         1&0&0&0& \\
         0&1&0&0& \\
         0&0&\rho^{-1}&0& \\
       0&0&0&\rho^{-1}& \\
\end{pmatrix}
\textsf{\textbf{M}}(1)
\begin{pmatrix}
         \rho^{3}&0&0&0& \\
         0&\rho^{-3}&0&0& \\
         0&0&\rho^{5}&0& \\
       0&0&0&\rho^{-5}& \\
\end{pmatrix}\nonumber
\end{equation}
and
\begin{eqnarray}\label{}
\mathbf{V}(\rho)= &&\frac{1}{4(1-2\nu)(1-\nu)}
\begin{pmatrix}
         2(1-\nu)\rho^{-1}\\
         -(1-2\nu)\rho^{-1} \\
         -(3-2\nu)\rho^{-2} \\
         -4(1-2\nu)\rho^{-2} \\
\end{pmatrix}\nonumber
\end{eqnarray}
in the shell, and $\mathbf{V}(\rho)= 0$ in the core.

At the interface ($\rho=1$),
\begin{equation}\label{001interface}
\begin{pmatrix}
         g_r \\
        g_\theta\\
        g_{r\theta} \\
        g_{rr}\\
\end{pmatrix}
=\textsf{\textbf{M}}(1)
\begin{pmatrix}
         \alpha^c_3 \\
         0 \\
         \alpha^c_5 \\
         0\\
\end{pmatrix}
=\textsf{\textbf{M}}(1)
\begin{pmatrix}
         \alpha^s_3 \\
         \alpha^s_{-3} \\
         \alpha^s_5 \\
         \alpha^s_{-5} \\
\end{pmatrix}
+ \mathbf{V}(1)
\end{equation}
and at the surface ($\rho=1/\sqrt{\eta}$)
\begin{equation}\label{}
\begin{pmatrix}
         g_r \\
        g_\theta\\
        g_{r\theta} \\
        g_{rr}\\
\end{pmatrix}
=\textsf{\textbf{M}}\left(\frac{1}{\sqrt{\eta}}\right)
\begin{pmatrix}
         \alpha^s_3 \\
         \alpha^s_{-3} \\
         \alpha^s_5 \\
         \alpha^s_{-5} \\
\end{pmatrix}
+ \mathbf{V}\left(\frac{1}{\sqrt{\eta}}\right)\nonumber
\end{equation}
The right-hand side can be written, using eq.~\ref{001interface}
\begin{equation}\label{001solution}
\textsf{\textbf{M}}\left(\frac{1}{\sqrt{\eta}}\right)
\begin{pmatrix}
         \alpha^c_3 \\
         0 \\
         \alpha^c_5 \\
         0\\
\end{pmatrix}
+ \mathbf{V}\left(\frac{1}{\sqrt{\eta}}\right)
-\textsf{\textbf{M}}\left(\frac{1}{\sqrt{\eta}}\right)
\textsf{\textbf{M}}^{-1}(1) \mathbf{V}(1)
\end{equation}
The condition that and the stress at the interface vanishes implies
that $\alpha^s_3$ and $\alpha^s_5$ are determined by equating the
last two lines of eq.~\ref{001solution} to zero.

The final result is:
\begin{eqnarray} \label{001alphas}
  \alpha^c_3 &=& \frac{-\nu+\eta^2[11-2\nu-20\eta+3\eta^2(3+\nu)]}{12(1-\nu)(1-2\nu)}\nonumber\\
  \alpha^c_5 &=& \frac{1-\eta^3[14-4\nu-25\eta+4\eta^2(3+\nu)]}{40(1-\nu)(1-2\nu)} \nonumber\\
  \alpha^s_3 &=& \frac{\eta^2[11-2\nu-20\eta+3\eta^2(3+\nu)]}{12(1-\nu)(1-2\nu)} \nonumber\\
  \alpha^s_{-3} &=& \frac{-5}{24(1-\nu)(1-2\nu)} \nonumber\\
  \alpha^s_5 &=& \frac{-\eta^3[14-4\nu-25\eta+4\eta^2(3+\nu)]}{40(1-\nu)(1-2\nu)}\nonumber\\
  \alpha^s_{-5} &=& \frac{3+\nu}{20(1-\nu)(1-2\nu)}
\end{eqnarray}
Here we have assumed that the stiffness constants are the same in
the core and in the shell. Different values of the stiffness
constants can be accommodated by writing different matrices
$\textsf{\textbf{M}}^c$ and $\textsf{\textbf{M}}^s$. And of course
this transfer matrix method can be extended to multishell NWs.
%

%
% BibTeX users please use
% \bibliographystyle{}
% \bibliography{}

\begin{thebibliography}{}

\bibitem{Wojn12} P. Wojnar, E. Janik, L. T. Baczewski, S. Kret, E. Dynowska, T.
Wojciechowski, J. Suffczynski, J. Papierska, P. Kossacki, G.
Karczewski, J. Kossut, and T. Wojtowicz, Nano. Lett. \textbf{12},
3404, (2012).

\bibitem {Niquet} Y. M. Niquet, C. Delerue and C. Krzeminski, Nano Lett. \textbf{12},
3545 (2012).

\bibitem {BoxbergNL} F. Boxberg, N. S{\o}ndergaard, and H.Q. Xu, Nano Lett. \textbf{10},
1108. (2010).

\bibitem {Amato} M. Amato, M. Palummo, and S. Ossicini, Mat. Sci. Engineering B \textbf{177}, 705
(2012).

\bibitem {Gutkin00} M. Yu. Gutkin, I. A. Ovid'ko and A. G. Sheinerman, J. Phys.: Cond. Mat. \textbf{12}, 5391 (2000).

\bibitem {Gronqvist} Johan Gr\"{o}nqvist, Niels S{\o}ndergaard  Fredrik Boxberg, Thomas Guhr, Sven {\AA}berg, and H.Q. Xu, J. Appl. Phys. \textbf{106}, 53508
(2009).

\bibitem{Maranganti}    R. Maranganti and P. Sharma, J. Comput. Theor. Nanosci. \textbf{4}, 715 (2007).

\bibitem {Nye}  J. F. Nye, Physical Properties of Crystals: Their Representation by Tensors and Matrices, Oxford University
Press, Oxford 1985.

\bibitem{Tsukrov}   I. Tsukrov and B. Drach, Int. J. of Solids and
Structures \textbf{47}, 25 (2010).

\bibitem {Marcus} P. M. Marcus and F. Jona, Phys. Rev. B \textbf{51} , 5263
(1995).

\bibitem {deformation} G. L. Bir and G. Pikus, Symmetry and Strain-Induced Effects in Semiconductors, Wiley, New York, 1974.

\bibitem {Aifantis} K. E. Aifantis, A. L. Kolesnikova and A. E. Romanov, Philosophical Magazine \textbf{87}, 4731 (2007).

\bibitem {Wortman} J. J. Wortman and R. A. Evans, J. Appl. Phys. \textbf{36}, 153
(1965).

\bibitem {Warwick} C. W. Warwick and T. W. Clyne, J. Mater. Sci. \textbf{26},
3817 (1991).

\bibitem{Eymery} Robert Koester, Jun-Seok Hwang, Damien Salomon, Xiaojun Chen, Catherine Bougerol,
Jean-Paul Barnes, Daniel Le Si Dang, Lorenzo Rigutti, Andres de Luna
Bugallo, Gwenoele Jacopin, Maria Tchernycheva, Christophe Durand,
and Joel Eymery, Nano Lett. \textbf{11}, 4839 (2011).

\bibitem {Davies} J. H. Davies, J. Appl. Phys. \textbf{84}, 1358 (1998).

\bibitem {Herstroffer} K. Herstroffer, R. Mata, D. Camacho, C. Leclere, G. Tourbot, Y. M. Niquet, A. Cros, C. Bougerol, H. Renevier and B. Daudin, Nanotechnology \textbf{21}, 415702
(2010).

\bibitem {Rigutti} L. Rigutti, G. Jacopin, L. Largeau, E. Galopin, A. De Luna Bugallo, F. H. Julien, J.-C. Harmand, F. Glas, and M. Tchernycheva, Phys. Rev. B \textbf{83}, 155320
(2011).

\bibitem {Polian} Polian, A., M. Grimsditch, I. Grzegory, J. Appl. Phys. \textbf{79}, 3343 (1996).

\bibitem {Wright} Wright, A.F., J. Appl.
Phys. \textbf{82}, 2833 (1997).

\bibitem {McNeil} McNeil, L.E, Grimsditch M., French R.H., J.
Am. Ceram. Soc. \textbf{76}, 1132 (1993).

\bibitem {Azuhata} T. Azuhata, M. Takesada, T. Yagi, A. Shikanai, S. F. Chichibu, K.
Torii, A. Nakamura, T. Sota, G. Cantwell, D. B. Eason, and C. W.
Litton, J. Appl. Phys. \textbf{94}, 968 (2003).

\bibitem {Kim} Young-Il Kim , Katharine Page  and Ram Seshadri, Appl. Phys. Lett. \textbf{90} , 101904
(2007).

\bibitem {Andre} R.Andr\'{e}, J.Cibert, Le Si Dang, J.Zeman, and
M.Zigone, Phys. Rev. B \textbf{53}, 6951 (1996).

\bibitem {Bester} Annie Beya-Wakata, Pierre-Yves Prodhomme, and Gabriel Bester, Phys. Rev. B \textbf{84}, 195207
(2011).

\bibitem {Boxberg} F. Boxberg, N. Sondergaard and H. Q. Xu, Adv.
Mater. 24, 4692 (2012).

\bibitem {Schulz} S. Schulz, M. A. Caro, E. P. O'Reilly and O.
Marquardt, Phys. Rev. B \textbf{84}, 125312 (2011).

\bibitem {Hocevar} Mo\"{\i}ra Hocevar, Le Thuy Thanh Giang, Rudeesun Songmuang, Martien den Hertog, Lucien Besombes, Jo\"{e}l Bleuse, Yann-Michel Niquet, and Nikos T.
Pelekanos,  Appl. Phys. Lett. \textbf{102}, 191103 (2013).

\bibitem {Adachi} S.Adachi, J. Appl. Phys. \textbf{58}, R1 (1985).

\bibitem {Vurgaftman} I. Vurgaftman, J. R. Meyer, and L. R. Ram-Mohan, J. Appl. Phys. \textbf{89}, 5815 (2001).

\bibitem {Montazeri} Mohammad Montazeri, Melodie Fickenscher, Leigh M. Smith, Howard E. Jackson, Jan Yarrison-Rice, Jung Hyun Kang, Qiang Gao, H. Hoe Tan, Chennupati Jagadish, Yanan Guo, Jin Zou, Mats-Erik Pistol and Craig E. Pryor, Nano Lett. \textbf{10}, 880
(2010).

\bibitem{Artioli} A. Artioli, P. Rueda-Fonseca, P. Stepanov, E. Bellet-Amalric, M. Den
Hertog, C. Bougerol, Y. Genuist, F. Donatini, R. Andr\'{e}, G.
Nogues, K. Kheng, S. Tatarenko, D. Ferrand, and J. Cibert, Appl.
Phys. Lett. \textbf{103}, 222106 (2013).

\bibitem {Wardzynski} W. Wardzynski, W. Giriat, H. Szymczak, and R. Kowalczyk, Physica
Status Solidi B \textbf{49}, 71 (1972).

\bibitem{LeSi89} Le Si Dang, J. Cibert, Y. Gobil, K. Saminadayar, and S. Tatarenko, Appl. Phys. Lett. \textbf{55}, 235
(1989).

\bibitem{Berlincourt} D.~Berlincourt, H.~Jaffe, and L.~R.~Shiozawa, Phys. Rev. \textbf{129}, 1009 (1963).

\bibitem{Hartmann} J.~M.~Hartmann, J.~Cibert, F.~Kany, H.~Mariette, M.~Charleux, P.~Alleyson,
R.~Langer, and G.~Feuillet J. Appl. Phys. \textbf{80}, 6257 (1996).

\bibitem{Pistol} M.-E. Pistol and  C. E. Pryor, Phys. Rev. B \textbf{78}, 115319
(2008).

\bibitem{Altarelli} M. Altarelli and N. O.
Lipari, Phys. Rev. B \textbf{15}, 4898 (1977).

\bibitem{Fishman}  Guy Fishman,
Semi-Conducteurs, les Bases de la Th\'{e}orie k.p, Les Editions de
l'Ecole Polytechnique, Paris 2010.

\bibitem{Trammel} Thomas E. Trammell, Xi Zhang, Yulan Li, Long-Qing Chen, Elizabeth C. Dickey, J. Cryst. Growth \textbf{310}, 3084
(2008).

\bibitem{Martin} R. M. Martin, Phys. Rev. B \textbf{6}, 4546 (1972).

\bibitem{Faux} D. A. Faux, J. R. Downes and E. P. O'Reilly, J. Appl. Phys. \textbf{82}, 3754 (1997).

\bibitem{Zhang} Lijun Zhang, Mayeul d'Avezac, Jun-Wei Luo, and Alex Zunger,
Nano Lett. \textbf{12}, 984 (2012).

\bibitem{Schmidt} V. Schmidt, P. C. McIntyre, and U. G\"{o}sele, Phys. Rev. B \textbf{77}, 235302
(2008).


\end{thebibliography}

\end{document}